# Relating the dynamics of photo de-mixing in mixed bromide-iodide perovskites to ionic and electronic transport

Ya-Ru Wang,[1,6] Marko Mladenović,[2,4] Eugene Kotomin,[1,5] Kersten Hahn,[1] Jaehyun Lee,[1] Wilfried Sigle,[1] Jovana V. Milić,[3] Peter A. van Aken,[1] Ursula Rothlisberger,[2] Michael Grätzel,[1,3] Davide Moia,[1,*] Joachim Maier[1,*]

[1]Max Planck Institute for Solid State Research, Stuttgart, Germany.

[2]Laboratory of Computational Chemistry and Biochemistry, Institute of Chemical Sciences and Engineering, École Polytechnique Fédérale de Lausanne (EPFL), Lausanne, Switzerland.

[3] Laboratory of Photonics and Interfaces, École Polytechnique Fédérale de Lausanne (EPFL), Lausanne, Switzerland.

[4] Integrated Systems Laboratory, Department of Information Technology and Electrical Engineering, ETH Zurich, Zurich, Switzerland

[5] Institute of solid-state physics, University of Latvia, LV 1063 Riga, Latvia

[6] Technische Universität München, München, Germany.

*moia.davide@gmail.com, office-maier@fkf.mpg.de

**ABSTRACT:** The observation of reversible de-mixing phenomena in mixed-halide perovskites under illumination is one of the most challenging as well as intriguing aspects of this class of materials. On the one hand, it poses critical constraints to the compositional space that allows reliable design of absorbers for perovskite photovoltaics. On the other hand, it holds potential for the development of novel optoionic devices where an ionic response is triggered via optical stimuli. Fundamental questions about the origin of such photo de-mixing process remain unanswered, both in terms of its mechanism as well as thermodynamic description. Here, we relate in-situ measurements of ionic and electronic transport of mixed bromide-iodide perovskite thin films performed during photo de-mixing with the evolution of their optical and morphological properties. The results point to the definition of different stages of the de-mixing process which, based on microscopy and spectroscopic measurements, we assign to regimes of spinodal decomposition and nucleation of quasi-equilibrium iodide- and bromide-rich phases. Combined with density functional theory calculations, we explore the role of dimensionality in the mechanism and reversibility of photo de-mixing and dark re-mixing processes, referring to electronic and ionic contributions to the de-mixing driving force. Additionally, our data emphasizes the role of the surface, as significantly different de-mixing dynamics, in terms of extent and reversibility, are observed for films with or without encapsulation. Our comprehensive analysis of transport, phase and optical properties of mixed-halide perovskites provides guidelines for future materials design as well as for the more general fundamental understanding of light-induced ionic phenomena.

## INTRODUCTION

Mixed bromide-iodide perovskites present great potential for optoelectronic devices due to the facile engineering of their properties.[1] One of the major challenges in the development of this class of materials relates to the process of photo de-mixing.[2, 3] Unlike the traditional concept of de-mixing, which is dictated by thermodynamic variables describing the equilibrium state[4] such as temperature and composition, photo de-mixing in mixed halide perovskites is a phase separation process that is triggered by light and that can be reversed in the dark (dark re-mixing).[2, 5] The (at least partially) reversible switching between the mixed phase, such as the 3D bromide-iodide perovskite MAPb($I_{0.5}Br_{0.5}$)$_3$ (MA: methylammonium cation ($CH_3NH_3^+$)), and the separated phases (iodide-rich and bromide-rich) via light/dark modulation indicates that light can reversibly vary the quasi-equilibrium thermodynamic phase properties of the system.[5, 6, 7] While it is widely accepted that the phase transformation is triggered by the increase in electronic charge carriers due to illumination or applied voltage bias, the precise nature of its driving force and of the relevant mechanisms mediating the process is still matter of debate.[2, 8] Importantly, the resulting phase heterogeneity in mixed halide perovskites on photo de-mixing causes an undesirable drop in power conversion efficiency of solar cells based on these materials, making the understanding of photo de-mixing a critical priority to advance the field.[9] More in general, controlling photo de-mixing and harnessing its potential represents both a compelling challenge and an opportunity for the emerging field of perovskite-based technologies.

From a thermodynamic point of view, the energy gain associated with photo-generated electronic charge carriers, especially holes, being "funneled" to I-rich phases from the mixed bromide-iodide phase has been discussed as a key contribution to the overall driving force for photo de-mixing.[10-13] In addition, localization of these carriers, as well as their interaction with the lattice or with specific ionic defects, have also been proposed as possible contributions to such driving force.[12, 14, 15] As it is clear that both electronic and ionic effects play important roles in the photo de-mixing and re-mixing of mixed-halide perovskites, improved understanding of both electronic and ionic charge transport properties during these processes would benefit the clarification of their origin and mechanism. In single halide perovskites (MAPbI$_3$), Kim *et al.* provided evidence for enhanced ionic response upon illumination, possibly related with an increase in iodide defect concentration.[7, 15] Later work focused on the discrepancies in defect formation pathways and light effects between iodide and bromide perovskites, pointing to an ionic contribution to the driving force for photo de-mixing. Specifically, the formation of iodide-rich domains may be required to allow for photo-induced high-order defect formation that stabilizes the overall structure.[15-17] Motti *et al.* reported that photo de-mixing in mixed halide perovskite films (MAPb(I$_{0.5}$Br$_{0.5}$)$_3$) causes negligible changes to the THz effective charge-carrier mobilities and therefore an absence of strong charge-carrier localization within the narrow-bandgap I-rich domain.[18] Despite these efforts, a direct measurement of the long-range electronic and ionic charge carriers' transport properties during photo de-mixing and correlation with the material's phase stability is still missing.

In this context, the role of the mixed halide perovskite composition on the thermodynamics and kinetics of photo de-mixing is a long-standing research question. In particular, halide perovskite compounds with different dimensionalities have been shown to undergo photo-induced phase separation with differences in the rate and extent of the phase instability.[5, 19] Of particular interest are the compounds where confinement occurs in 1 dimension (two-dimensional (2D) halide perovskites, A'A$_{n-1}$B$X_{3n+1}$, n is the number of perovskite layers between each layer of spacer molecules) and three dimensions (3D nanocrystals (NC), A''-ABX$_3$, where A'' is a ligand that functionalizes the crystal surface).[20] Recent reports on 2D compounds have revealed a slower rate of the photo de-mixing process in these systems compared with their 3D counterpart,[21] which is consistent with the low electronic and ionic conductivities for the 2D compounds.[22] Recent reports have also shown that the nature of the spacer used in the 2D structure is a determining factor regarding the occurrence and kinetics of photo de-mixing.[21, 23] The role of photo stability of 2D bromide perovskites has also been found to play an important role in the occurrence of photo de-mixing in 2D mixed bromide-iodide perovskites.[24] Interestingly, enhancement of ion transport occurring under light also in 2D systems has been suggested.[25] On the other hand, it has been demonstrated that NC mixed halide perovskites show limited phase instability,[26] an aspect that can be traced back to the contribution of the additional free energy cost of forming an interface.[27] Notably, the use of 2D and NC halide perovskites has, in several instances, been associated with improved stability and performance in devices. Such trend has often been discussed in terms of the "ion blocking" properties of the spacers or ligand molecules used in these systems, although a quantitative picture supporting such explanation is missing.[38-4028] These aspects further emphasize the importance of extending the investigation of photo de-mixing in mixed halide perovskites across different dimensionalities, and in relation to their charge transport properties.

The difference in the photo de-mixing process for compounds with different dimensionalities suggests significant changes in the free energy landscape describing their phase stability due to variations in structure and composition. Investigation of this aspect through microscopy methods that allow probing the morphology or phase properties (i.e. spatial distribution and domain size) can be helpful in the identification of the de-mixing mechanism. However, to date, this has been challenging, mostly due to the instability of the more heavily investigated 3D perovskite thin films. An early report by Bischak et al. demonstrated that low photon energy emitting regions detected with cathodoluminescence, identified as iodide-rich domains, are localized in the proximity of the grain boundaries in thin films of MAPb(I$_{0.5}$Br$_{0.5}$)$_3$ upon illumination.[12] Mao et al. observed the photoluminescence emission of nanocluster-like I-rich domains not only at the edges of MAPb(Br$_x$I$_{1-x}$)$_3$ microplatelets but also throughout their bulk.[29] However, direct measurements of the domains' composition and size are still missing.

In this work, the phase behavior and charge transport properties of mixed bromide-iodide perovskites with different dimensionality (2D, 3D, and nanocrystals) under light are investigated. Firstly, we focus on 2D Dion-Jacobson mixed bromide-iodide perovskite thin films with composition (PDMA)Pb(Br$_{0.5}$I$_{0.5}$)$_4$ (PDMA: 1,4-phenylenedimethanammonium spacer) as a model system to study photo de-mixing, due to its established reversibility.[5] The phase behavior of the mixture is investigated both under light and in the dark using a wide range of in-situ and ex-situ experimental techniques as well as theoretical calculations. The conductivity changes during de-mixing are monitored by impedance measurements in the dark and under light. This allows us to probe the ionic and electronic transport properties of the thin films, and to correlate them with the changes in phase and morphological properties during photo de-mixing and dark re-mixing, which are tracked via time-dependent in-situ optical absorption spectroscopy and ex-situ microscopy techniques. Next, the dependence of the phase behavior of mixed halide perovskites on dimensionality (3D, and nanocrystals) is discussed, based on the comparison of their optical absorption and electrical response under similar experimental conditions used for the 2D systems. Additional insights are given on the role of surface encapsulation of the films in their photo de-mixing and dark re-mixing. Finally, with the aid of Density Functional Theory (DFT) calculations, we discuss under which conditions the role of not only electronic but also ionic effects should be accounted for in the evaluation of the driving force of photo de-mixing in the 3D and 2D mixed bromide-iodide perovskites investigated here.

## RESULTS AND DISCUSSION

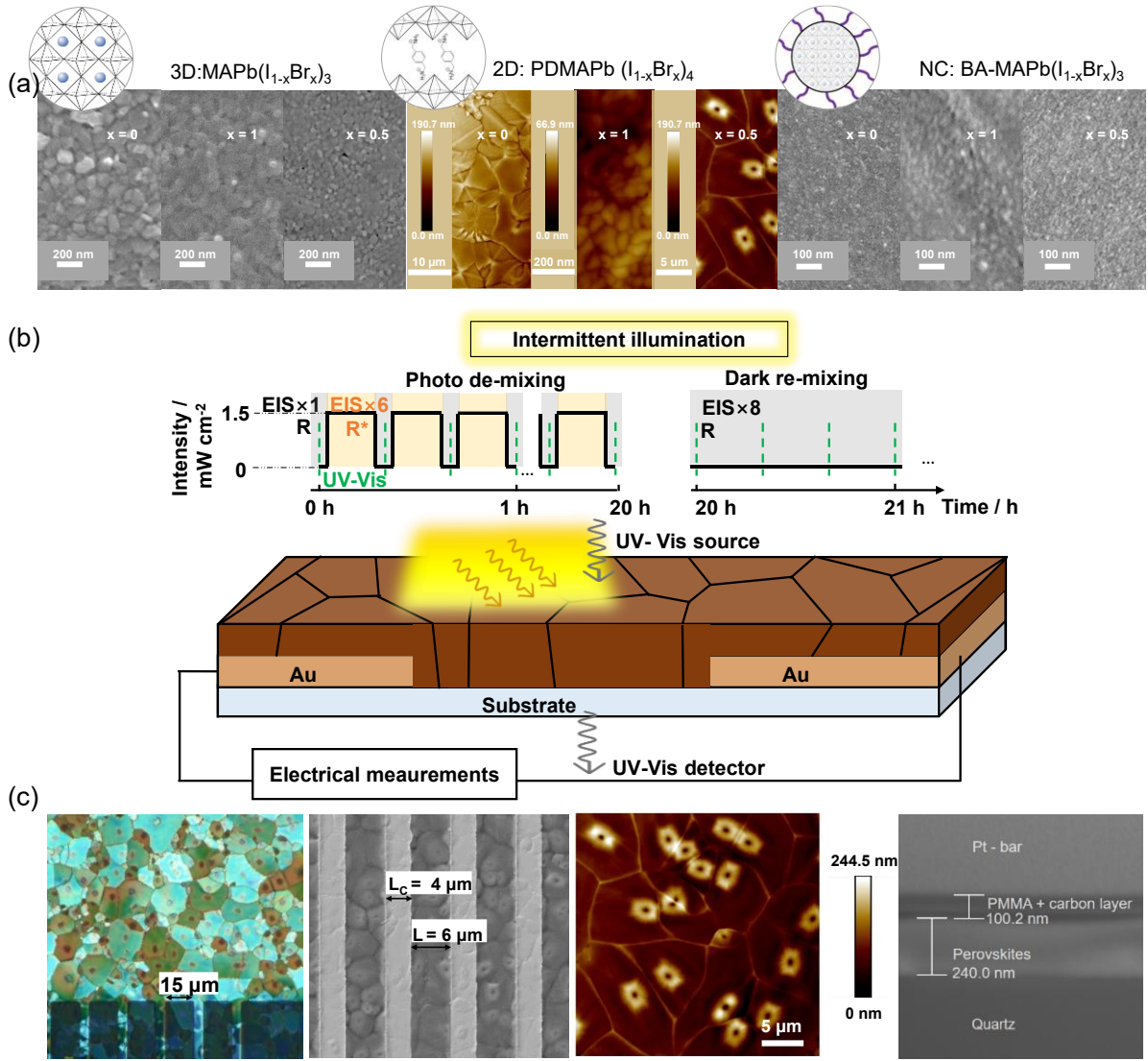

**Figure 1 Top view morphology of halide perovskites with different dimensionalities:** (left) Top view SEM images of 3D MAPb($I_{1-x}Br_x$)$_3$ thin films; (center) Top view AFM images of 2D (PDMA)Pb($I_{1-x}Br_x$)$_4$ thin films and (right) Top view SEM images of nanocrystalline (NC) BA-MAPb($I_{1-x}Br_x$)$_3$ thin films. MA, PDMA and BA refer to methyl ammonium (MA$^+$), 1,4-phenylenedimethanammonium (PDMA$^{2+}$) and n-butylammonium (BA$^+$). (b) Schematic representation for simultaneous electrical and optical characterization of a 2D (PDMA)Pb($I_{0.5}Br_{0.5}$)$_4$ thin film during photo de-mixing and dark re-mixing. Modulation of the bias light during photo de-mixing (300 s dark, 900 s light) allows for UV-Vis (in the dark) and impedance (EIS) measurements (conducted every 150 s, both in the dark and under light). UV-vis and EIS measurements are recorded also after switching off the light during the dark re-mixing process. (c) Top view images of a (PDMA)Pb($I_{0.5}Br_{0.5}$)$_4$ thin film in the pristine state on a quartz substrate with interdigitated gold electrodes. From left to right: optical microscope image (polarized light), SEM image, AFM image (topography) and cross-sectional SEM image of a sample encapsulated with PMMA.

The thin film morphology and structure (inset) of the mixed halide perovskite compounds and their end members with different dimensionalities investigated in this study are shown in Figure 1a. The mean grain aggregates size for the 3D, 2D and NC mixed halide perovskites is around 150-200 nm, 3-5 µm, 10 nm respectively. The XRD patterns and UV–Vis spectra confirmed the successful preparation of such mixed halide perovskite thin films (Figure S2). A distinct feature of 2D mixed halide perovskites is the appearance of the XRD peak at a very low angle, which corresponds to the distance of the lead halide octahedral slabs.[5] Compared with 3D mixed halide perovskites, the NC-based film exhibits absorption at slightly higher photon energies due to quantum confinement effects. Such effects have an even greater influence on the absorption spectrum of 2D mixed halide perovskites investigated here, which shows a sharp excitonic peak around 450 nm.[30]

**I. Photo de-mixing and dark re-mixing in 2D mixed halide perovskites**

We consider a thin film (thickness 240 nm, Figure 1c, right, SEM) of the 2D mixed-halide perovskite (PDMA)Pb(I$_{0.5}$Br$_{0.5}$)$_4$ encapsulated with a thin film (~ 100 nm) of PMMA (Poly(methyl 2-methylpropenoate)). A similar morphology is found when the thin films are deposited on a bare quartz substrate or on a quartz substrate with interdigitated electrodes used for conductivity measurements. This allows the discussion of electrochemical measurements on substrates with electrodes also in relation to the analysis of the films' morphological properties obtained using substrates without electrodes.

The resistance measured within the periodic dark time slots during the photo de-mixing (R, upper panel in Figure 2b) decreases abruptly by more than one order of magnitude under illumination within the first 20 minutes. From the data, it is not possible to comment on whether such decrease in resistance occurs gradually within this short time range or immediately after exposure to light. Further illumination leads to a monotonic increase in this resistance, which however remains lower than the initial value before illumination within the 20-hour illumination time. The electronic resistance under light (R*, lower panel in Figure 2b) decreases abruptly within the first 20 minutes of illumination, it then remains approximately stable for approximately 1 h, before decreasing slowly to a lower value with a time constant of approximately 1.6 h. A relatively stable value with only a slight increase in R* is observed in the 10–20 hour time range. Based on these data, two distinct stages seem to describe the mechanism for photo de-mixing in the 2D perovskite (PDMA)Pb(I$_{0.5}$Br$_{0.5}$)$_4$. While R* can safely be attributed to an electronic contribution (see DC polarization in Figure S8), the interpretation of R is more complex and will be addressed below. Figure 2c (bottom panel) shows the change in absorbance (ΔA) during photo de-mixing. The data highlights bleaching of the absorption feature associated with the excitonic peak of the pristine phase and the emergence of two features at short and long wavelengths. Following previous interpretations,[5, 19] these changes in absorption features indicate that under illumination the single-phase mixed halide perovskite separates into Br-rich and I-rich phases. It is worth noting that within the first 20 minutes of illumination, an absorption feature at long wavelength associated with the iodide-rich phase emerges with a peak in ΔA around 470 nm (Figure 2c, bottom left). With continuous illumination, an absorption feature at longer wavelengths (~510 nm) emerges for the I-rich phase, which becomes increasingly enhanced until quasi-equilibrium is reached. On the other hand, the spectral shape of the absorption changes associated with the formation of the Br-rich phase does not vary significantly throughout the process (Figure 2c, bottom right).

Consistent with the electrical characterization, the optical absorption results suggest that photo de-mixing occurs through a two-stage process. During a 1$^{st}$ stage (first ~20 minutes under illumination), composition broadening close to the initial composition of the mixture occurs, a process that could be due to spinodal decomposition of the compound (see below). During a 2$^{nd}$ stage, formation and subsequent growth of the quasi-equilibrium phases occur. Previous work has presented evidence highlighting such multi-stage nature of the photo de-mixing process.[31] The observed decrease in R during the 1$^{st}$ stage could be related with an increase in ionic conductivity, which would occur as a result of the formation of mobile charged point ionic defects along with possible formation of high-order defects involving trapping of the electronic carriers.[16] Slow (~minutes timescale) de-trapping of trapped electronic charge carriers could also lead to the observed decrease in R. Finally, despite encapsulation of the sample, other effects associated with stoichiometry changes due to illumination and consequent variation in the equilibrium electronic charge carrier concentrations in the film cannot be ruled out.[32] The change in electronic resistance under light (R*) during the 1$^{st}$ stage is potentially related with an increase in the concentration of electronic carriers due to the formation of the I-rich clusters, which widens the optical absorption window of the film. Such change in the photo-generation rate associated with the composition broadening close to the initial composition (1$^{st}$ stage) and with the final I-rich phases' compositions (2$^{nd}$ stage) correlate with the overall trend of R* as a function of time under light. Additional effects are expected to play a role. For example, the increased phase disorder in the film as function of illumination time reduces percolation pathways for electrons and holes. These are funneled to the I-rich domains, where they can either percolate to the contacts or remain trapped and recombine. Changes in recombination rates and in long-range mobility of the carriers may be responsible for the complex shape of R* vs time. The morphology or phase properties changes during photo de-mixing can also alter the pathways for electronic and ionic transport, therefore influencing the electrical and optical response.

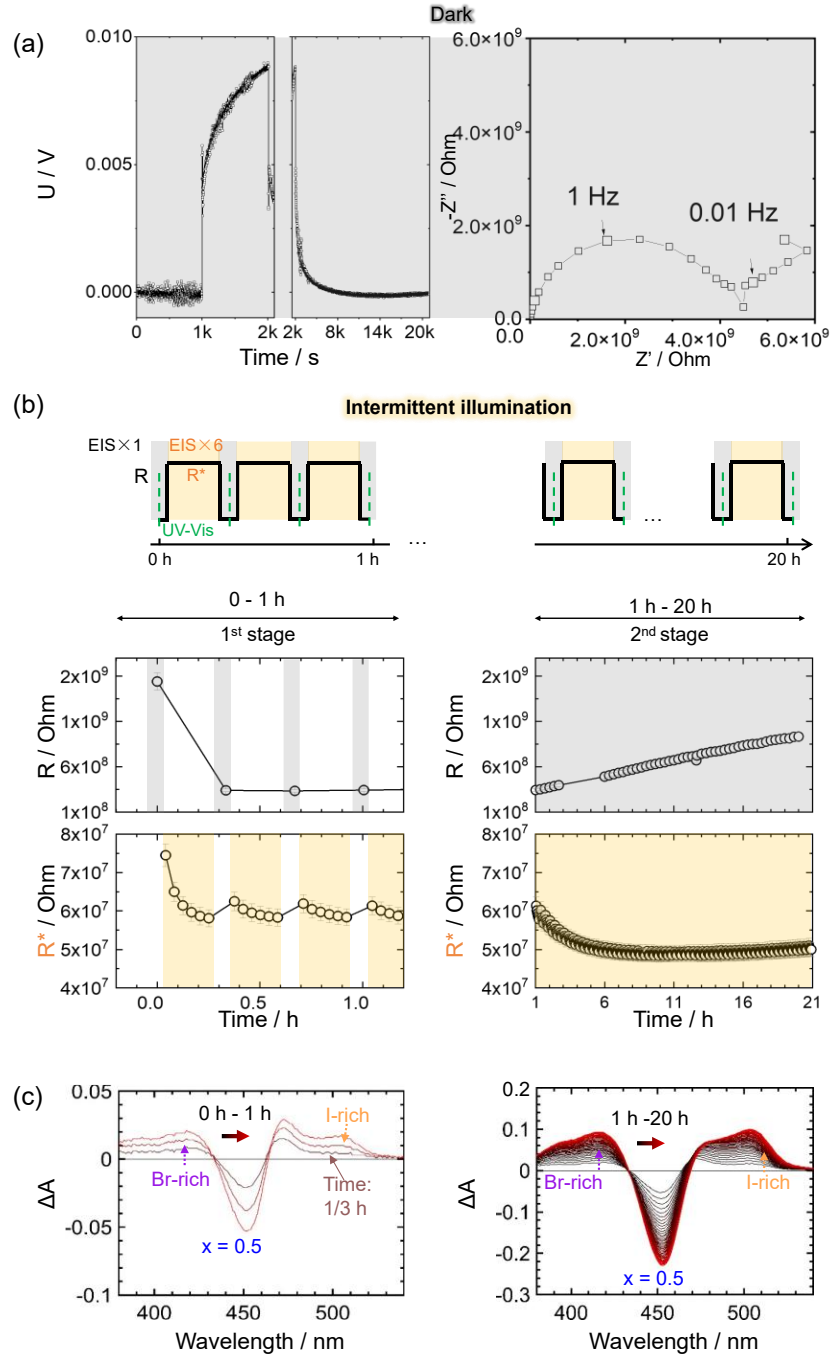

**Figure 2. Electrical and optical characterization** of (PDMA)Pb(I$_{0.5}$Br$_{0.5}$)$_4$ thin films (with PMMA encapsulation and in Ar atmosphere) (a) Electrical measurements performed in the dark close to equilibrium. Partial electronic and ionic conductivities ($\sigma_{eon}$ and $\sigma_{ion}$) are extracted from galvanostatic and impedance measurements. (b) Resistance and (c) optical absorbance evolution under intermittent illumination condition at 80 °C for a (PDMA)Pb(I$_{0.5}$Br$_{0.5}$)$_4$ thin film. The light ON/OFF conditions are modulated such that there are three complete cycles per hour (150 s dark, 900 s light, 150 s dark, See Supporting information in section S4.1 for more details). For each cycle, UV-Vis measurements are taken when the samples are in the dark. Impedance measurements are conducted 2 times in the dark (grey background) and 6 times under light (1.5 mW / cm$^2$, yellow background). During this process, the evolution of the resistance (upper panels) and UV-Vis spectra (change in absorbance, ΔA) are extracted (2c). Two stages of de-mixing are identified: ~1 h (left) and ~1–20 h (right). Both values of the resistance in the dark (R) and under light (R*, under 1.5 mW / cm$^2$) are tracked during photo de-mixing.

To access information on the changes in morphology and phase properties of the perovskite surface pre- and post-illumination, three sets of SEM images were carried out on the same area of one PDMAPb(I$_{0.5}$Br$_{0.5}$)$_4$ thin film deposited on the interdigitated gold electrodes (See Supporting information in section S5.1, Figure S12-16). SEM measurements were conducted on the same position, when the sample is in its pristine state, after 20 h illumination, and after being annealed at 80 °C for 30 days in Ar. After 20 h illumination, small white domains are observed close to the grain boundaries. Surprisingly, in the area where they were heavily exposed to the E-beam, needle-like structure appeared (Figure S15), revealing the changes within the perovskite thin films when being exposed to electron beam. To minimize the influence of E-beam, ex-situ SEM measurements for these 2D thin films without encapsulation are carried out. Silicon substrates with a thin top layer of $Si_3N_4$ and with small windows where the silicon is removed are used (Figure S17). This choice allows one to perform SEM measurements with good resolution without conductive coating and it also offers the possibility of conducting imaging measurements in transmission mode. To study the changes in perovskite thin film morphology during photo de-mixing, the illumination time was varied (0 h, 1 h, 20 h, 50 h) as shown in Figure 3a (a back scattered electron (BSE) detector is used, see Figures S18–S21 for measurements with other detectors). After 1 h illumination, no significant changes are observed compared with the pristine sample. However, after 20 h of illumination, small crystallites with a width of approximately 30 – 70 nm form close to the grain boundaries in the film. With 50 h illumination, the small crystallites form not only close to the grain boundaries but also inside the grains. Intriguingly, the crystallites increase in concentration with longer illumination time, but their size remains approximately constant. Figure S22 shows the morphology of the crystallites that are formed using transmission mode measurements (measurements of bulk) and further confirms these findings. More cracks are present after illumination in areas that are close to the grain boundaries. These could be due to the strain induced when new phases are formed. The formation of nano-domains for long enough illumination is qualitatively consistent with the two-stage description outlined above for the de-mixing process. The initial stage with largely unchanged morphology, potentially associated with spinodal decomposition, is followed by a second stage with nucleation of nanoscale domains.

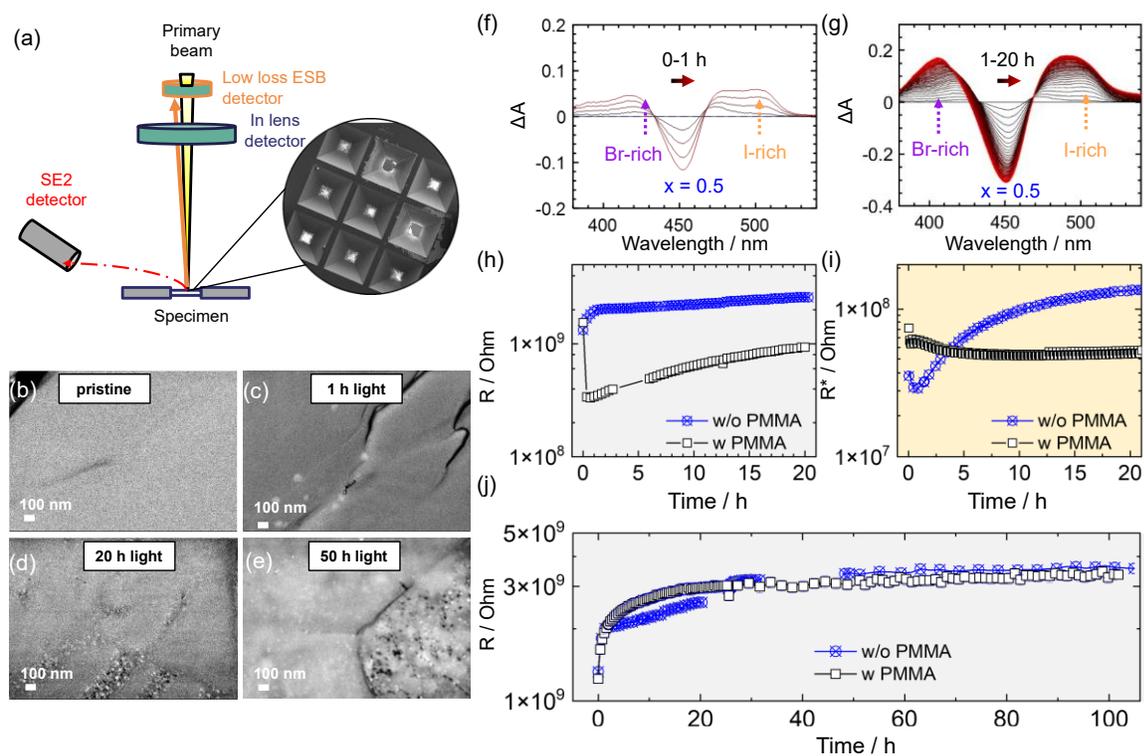

**Figure 3** (a) Schematic of the geometry of the three different detectors for SEM measurements applied to $(PDMA)Pb(I_{0.5}Br_{0.5})_4$ thin films (without encapsulation) coated on silicon supported $Si_3N_4$ (9 windows, 100 × 100 μm, 15 nm in thickness) substrates (see Figure S17 and 19). (b-e) SEM Images obtained using a BSE detector (sensitive to phases and compositions, magnification: 30k) with different photo de-mixing time: pristine (b), de-mixed for 1 h (c), 20 h (d), and 50 h (e). All experiments are performed at 80 °C, illuminated via white LED illumination intensity of 1.5 mW/cm². Complete SEM image sets with different detectors (SE2, BSE and Inlens) and magnifications (5K and 30 K) can be found in Figure S20–S21. Note that these measurements refer to 4 different samples from the same batch and exposed to the same illumination conditions for different time, to avoid possible influence on the materials' properties due to E-beam exposure (see Supporting Figure S12–16 for similar experiments conducted on one sample). (f-g) Change in absorbance (ΔA) evolution of $(PDMA)Pb(I_{0.5}Br_{0.5})_4$ thin films (without PMMA encapsulation and in Ar atmosphere) during photo de-mixing: (f) 1st stage, 0 – 1h (g) 2nd stage 1h – 20 h illumination. (h) Resistance in the dark (R, grey background) and (i) under light (R*, yellow background) obtained from fitting impedance data collected during photo de-mixing. (j) Resistance changes during dark re-mixing after photo de-mixing. The data compares a thin film without PMMA encapsulation (blue crossed circles) and one with PMMA (black square) encapsulation on the surface. All measurements in (f–j) are performed under Ar atmosphere.

The latter process occurs at early times (<10 hours) in proximity of the grain boundaries (heterogeneous nucleation), while at longer times (> 10 hours) it extends to the inner regions of the grains (heterogeneous or possibly even homogeneous nucleation). This indicates the presence of special sites close to the grain boundaries that are capable of significantly lowering the activation energy associated with the nucleation process.

To investigate the chemical nature of the crystallites detected with SEM after photo de-mixing further, cross-sectional TEM of a de-mixed film was conducted. Au-coated substrates were used here to decrease the charging effects. Figure S25a shows a top-view SEM image of a de-mixed film. Similar to the above-mentioned results, "bright" crystallites are observed close to the grain boundaries. One single crystallite is identified and investigated with further measurements (Figure S25b, c). Figure S25d and S25e show the TEM image of the cross section of such region in proximity of the grain boundary. EDX mapping evidenced vertical segregation, whereby iodide preferentially segregates close to the grain boundary groove at the substrate/perovskite interface with an average composition of $I_2Br$ while bromide segregates to the free perovskite surface with an average composition of $IBr_2$ (Figure S25e). Such stoichiometries are averaged over a lamellae thickness of 60 nm and therefore are not necessarily representative of the quasi-equilibrium composition associated with de-mixed domains. This experiment provides direct evidence that I-rich nanodomains form upon photo de-mixing of the investigated 2D mixed-halide perovskites, an observation that is qualitatively consistent with previous work on 3D systems.[12] Interestingly, with a longer illumination time of 80 h, different contrast of the formed crystallites suggests that formation of both I-rich and Br-rich domains may occur at a later stage of photo de-mixing (Figure S27). The EDX mapping of such a sample highlights a bromide-rich domain formed on the perovskite surface (Figure S28). The elementary analysis in Table S12 shows that the iodide to Pb ratio (I : Pb) and bromide to Pb ratio (Br : Pb) are 0.96 and 1.89 respectively, indicating that at a long illumination time, significant photo-induced degradation may occur simultaneously to the photo de-mixing process due to loss of $I_2$.

As the morphological investigation has been carried out for films without encapsulation, the effect of such parameter on photo de-mixing was evaluated by conducting the same experiment presented in Figure 2a and b on another film, but without adding the thin layer of PMMA after the deposition of the 2D mixture (see Figure S10 and S11 for optical and impedance data). Our previous report already emphasized the importance of encapsulation in the optical and structural reversibility of the photo de-mixing process.[5] Figure 3b shows the changes in absorption detected during photo de-mixing for the non-encapsulated sample. While in Figure 2a a clear change in spectral shape for the I-rich domain is observed during photo de-mixing in the beginning, the absorption data collected during photo de-mixing of the film without encapsulation suggest that the two phases' compositions are already close to the final compositions within the first hour of illumination. If we were to assign the long wavelength feature (at 510 nm) to the nucleated I-rich nanodomains, this would suggest that faster nucleation of such domains occurs without encapsulation of the thin film compared to films with PMMA. For illumination occurring over longer times, such absorption features grow to a lesser extent compared with the one reported in Figure 2b.

Figure 3h, i show the comparison of the resistance fitted to the impedance data collected during photo de-mixing for the dark and light cycles referring to the films with or without PMMA. The data show remarkable differences: contrary to the abrupt decrease in R within the first 20 min observed for the encapsulated sample, an increase in such resistance was observed for the film without PMMA encapsulation. Such distinct difference indicates that the properties of the surface play an important role in the determination of the electrical response of the film during photo de-mixing. It is worth noting that the value of R measured in the dark (the first point in Figure 3h) prior to illumination is very similar for both the encapsulated and the non-encapsulated case, and that differences in this quantity arise only upon illumination.

Concerning the electronic resistance under light (R*, Figure 3i), this shows a lower value at early time scales for the non-encapsulated sample compared with the encapsulated case. Interestingly, the value of R* for the sample without PMMA shows a decrease in resistance within the first hour followed by an increase as a function of de-mixing time, in contrast with the almost monotonically decreasing trend recorded for R* in the encapsulated film. Indeed, at long time scales (~ 4h), R* is larger for the sample without PMMA than for the sample with PMMA. Throughout the measurement, a transient in the value of resistance under light is observed (R* in yellow background) when the light is switched off and back on (Figure S31). Such transients indicate short time scale (~20 min) transients of R* that are consistent with the overall trend of R* shown in Figure 3i. These emphasize the tendency for R* to either increase or decrease under light following short dark intervals, depending on whether the film is encapsulated or not, respectively. Finally, while the negligible changes in absorption over the last hours of the experiment (Figure S7) would suggest that the photo de-mixing process leads to a state of quasi-equilibrium for both films with and without surface encapsulation within this time scale, the electrical data highlight changes in transport properties even after 20 hours of illumination at 80 °C.

Given the very thin layer of PMMA used here (~60 nm, See Figure S30) and its optical properties (largely transparent in the emission wavelength range of the white LED used here as bias light), we can exclude that the larger resistance R* observed for the encapsulated sample on illumination at early times is due to optical shadowing by such layer. In addition, PMMA is often used as passivation layer in the field of halide perovskites, generally leading to longer charge carrier lifetimes. The different trends observed for R* in these two cases may suggest that for the encapsulated film the increase in generation rate due to the increase in optical absorption at long wavelengths may dominate the impedance evolution. Additionally, a less favorable morphology for electron transport may lead to an increase in resistance for the sample without encapsulation. Importantly, XRD measurements performed before and after de-mixing show that no degradation products such as PbI$_2$ or PbBr$_2$ are observed (Figure S26), indicating that no significant photo-induced degradation occurs within this time scale also for samples without encapsulation. We hypothesize that the pronounced nucleation of nano-domains visible for the sample without PMMA occurs to a lesser extent when the surface is encapsulated, due to a lower surface energy associated with the perovskite/PMMA interface compared to the perovskite/gas interface. This influences the evolution of the absorption spectra of the two films, as well as of their photo-conductivity.

A possible explanation for such difference in behavior may be related with trapping effects which influence the time dependent charge carrier concentration in the film. Gradual trapping of photo-generated charges and trap-filling in the film would explain the gradual decrease in resistance under light. If such trapping were linked to a more complex mechanism, such as iodide oxidation and excorporation of iodine into the gas phase[7] according to

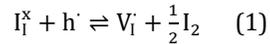

$$I_I^x + h^\cdot \rightleftharpoons V_I^\cdot + \frac{1}{2}I_2 \quad (1)$$

it would be possible to explain the observed trends further. Indeed reaction (1) is hindered when the film's surface is encapsulated with PMMA. In such case, after the formation of I-rich domains, such domains would be preserved, possibly explaining the sharper absorption feature at long wavelengths arising after long time illumination. Without encapsulation, the rate of the forward reaction in (1) would be enhanced by light, potentially competing with the formation of the I-rich domains (reduced long wavelength absorption feature in Figure 3a). Such excorporation process could also be associated with an increasingly defective surface of the films, which may lead to enhanced recombination of charge carriers and explain the observed increase in R*. Interestingly, earlier studies showing changes in PL intensity on de-mixing in mixed halide perovskite thin films indicated poor reversibility across dark-light cycles performed on films without encapsulation, an observation that could be correlated with our data.[33]

The change in resistance in the dark during re-mixing is displayed in Figure 3j. Without PMMA encapsulation, the re-mixing kinetics exhibit a slightly more complex sequence of steps compared with the encapsulated sample. Overall, the change in resistance and the time constant are nearly the same in both cases. This indicates that, as also observed for the electrical properties of the pristine films in the dark, surface encapsulation does not significantly affect the kinetics of dark re-mixing. The resistance measured after 100 hours in the dark is larger by a factor of two compared with the value recorded for the pristine case (prior to photo de-mixing and dark re-mixing), indicating that the changes in electrical properties due to illumination are not reversible within the length of this experiment. Importantly, this is the case also for the encapsulated sample. This suggests that, even though substantial reversibility is observed for the optical and structural properties of these 2D mixed halide perovskites with encapsulation already after 100 hours at 80 °C, the resistance of the sample after photo de-mixing and dark re-mixing does not fully recover within this timeframe, but instead it relaxes to a larger value than the starting one.

From a morphological point of view, the domains that appear along and close to the grain boundaries after photo de-mixing show a tendency to disappear in the dark (re-mixing performed in argon glovebox at room temperature for 6 months, Figure S32). This confirms that photo de-mixing induces morphological changes that are, at least partially, reversible.[5] Full reversibility from a morphological point of view is difficult to prove, due to the need of non-encapsulated surfaces for SEM measurements. Under such conditions, halogen exchange with the gas atmosphere during the experiment is expected to occur, inducing irreversible degradation.

**II. Role of dimensionality in photo de-mixing**

In this section, we address the effect of dimensionality on the evolution of optical and electrical properties during photo de-mixing and dark re-mixing of mixed-halide perovskite thin films. By using a similar approach as the one described above, we perform a simultaneous electrical and optical investigation of 3D MAPb($I_{0.5}Br_{0.5}$)$_3$ films and comment on the similarities and differences compared with the results obtained for the 2D perovskites.

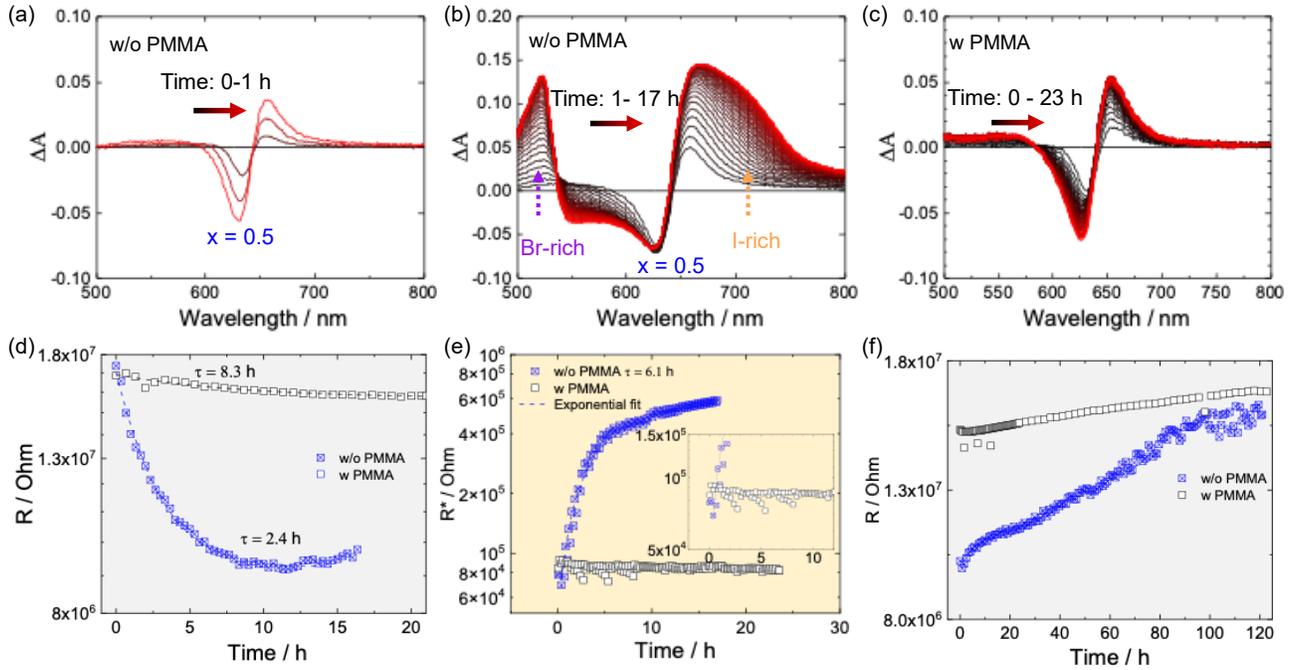

**Figure 4** (a - c) UV-Vis spectral evolution of MAPb($I_{0.5}Br_{0.5}$)$_3$ thin films (a-b) without PMMA encapsulation and in Ar atmosphere under light (white LED, 1.5 mW/cm$^2$) at 40 °C for 17 h. Measurements were performed with 20 min interval. See Figure 1a for details about the dark and light modulation procedure. (a) First stage of photo de-mixing: illumination induces composition fluctuation close to the initial composition; (b) second stage of photo de-mixing: formation of the two distinct absorption features that correspond to the final Br-rich and I-rich phases. (c) Photo de-mixing of a 3D film with PMMA encapsulation and in Ar atmosphere under light (1.5 mW/cm$^2$) at 40 °C for 23 h. Only the first stage of photo de-mixing is observed. (d - f) Kinetics of the change in resistance of MAPb($I_{0.5}Br_{0.5}$)$_3$ thin films with PMMA encapsulation (black square) and without encapsulation (blue crossed square) measured during intermittent illumination (d) in the dark (R) and (e) under light (R*), and (f) during dark re-mixing.

A two-stage photo de-mixing behavior is also observed for 3D mixed halide perovskites, when probing their optical properties under light (Figure 4a, b). At early time scales (~1h), the change in the absorption spectrum of the 3D MAPb($I_{0.5}Br_{0.5}$)$_3$ thin film on illumination indicates compositional fluctuation that is close to the initial composition (1$^{st}$ stage). Interestingly, for this case, we do not observe a clear emergence of an absorption feature at short wavelengths associated with the bromide-rich phase, as we report for the 2D sample. On the other hand, after one hour of continuous illumination, the formation of the two distinct absorption features, previously identified as I-rich and Br-rich phases, is detected (2$^{nd}$ stage).[2] Here the iodide-rich absorption feature gradually red-shifts in time, while a bromide-rich absorption feature emerges within a specific wavelength range, suggesting possible nucleation of this phase. This result already suggests that the process of photo de-mixing shares fundamental similarities, but also some differences when varying the dimensionality of mixed bromide-iodide perovskites from 2D to 3D.

The effect of surface encapsulation is evaluated by conducting photo de-mixing experiments under the same conditions but encapsulating the film with a PMMA layer on the surface, similarly to the 2D case. Interestingly, only the first stage of photo de-mixing is observed here (Figure 4c), while for long time illumination, under the same conditions as for the experiment in Figure 4a and b, no significant change in the absorption spectrum is detected. Once again, this result suggests that encapsulation plays an important role in the reaction pathway leading to the photo de-mixed state, also in 3D mixed halide perovskites. In particular, following the learning points from the previous section, it appears that the heterogenous nucleation characteristic of the second stage may be inhibited by encapsulation in the 3D system to an even larger extent than in the 2D film. Such effect would point to surface defects as possible players in terms of nucleation centers. Passivation of the surface with PMMA might reduce the reactivity of such centers, reducing the concentration of kinetically effective paths to the formation of the final phases. Changes in the thermodynamic properties in the system could also play a role. For example, reduction in electron-hole recombination due to surface passivation could lead to an increase in electronic charge concentrations under light, although this factor alone would point to a larger driving force for the encapsulated film. Previous work on the dependence of $I_2$ permeability on ionization energy of hole transport materials and insulating PMMA suggests that PMMA, with its deep HOMO energy, suppresses iodine migration and therefore the degradation mechanisms associated with iodine excorporation from the perovskite.[34] Therefore, whether the difference in behavior between the two samples with and without encapsulation is dictated by thermodynamic and/or kinetic arguments remains an open question.

For the encapsulated film, the photo de-mixed state shows reversible behavior when light is switched off (Figure S33). For the film without PMMA, however, only partial reversibility in the absorption feature is observed, with clear formation of $PbI_2$ (Figure S34, supporting information). This observation could be related to the relevance of reaction (1) for the film with exposed surface. Oxidation and subsequent excorporation of bromine following a similar mechanism may also be possible (more on this below).

To study the effect of encapsulation on the electrical response during photo de-mixing and to check the reversibility of the process, the impedance of the two samples discussed above is measured during illumination and after light has been switched off. The kinetics of the resistance changes in the dark and under light during photo de-mixing (gray and yellow background, respectively) and in the dark during re-mixing (gray background) under these two conditions (non-encapsulated vs encapsulated with PMMA) are displayed in Figure 4d, e, and f.

After 20 hours of photo de-mixing, the value of R decreases by only about 7% for the film with PMMA, consistent with the minor change in optical absorption. On the other hand, the value of R for the sample without PMMA, decreases by a factor of two under similar conditions. Interestingly, both the change in resistance and the change in absorption recorded after 1h for the sample without encapsulation (completion of the first stage) are approximately the same as the values recorded for the encapsulated sample after 23 h illumination. This indicates that there is good correspondence between the (dark) electrical and the optical properties of the two films, and that the two samples reach similar states, on completion of the first stage of photo de-mixing. The data associated with the non-encapsulated mixed halide perovskite film in Figure 4c show that no distinct separation of the two stages (as seen in the optical absorption evolution) is observable from this electrical response in the dark; the value of R decreases monotonically for the first 10 hours, before undergoing a slight increase. Such behavior is likely to be due to the combined effect of increased charge carrier concentration (in the dark) within the growing, gradually smaller bandgap, iodide rich regions and the increasingly unfavorable phase composition that hinders charge transport as the de-mixing process progresses. Possible trapping effects or stoichiometry changes that increase the electronic charge concentration in the dark after exposure to light may also play a role, as discussed for the 2D films.

The resistance under light, R*, undergoes an increase of almost one order of magnitude for the non-encapsulated film. This would point to the lack of percolation for the I-rich phases that form in the film and that the rate of recombination increases the more the de-mixing advances. The degradation of the film and formation of $PbI_2$ could also create insulating barriers that impede charge percolation. Strikingly, for the film with PMMA encapsulation, the electronic resistance under light remains almost constant with respect to time during the whole photo de-mixing process. This difference may be related to the protective role of the PMMA encapsulation layer, which reduces degradation of the film and the excorporation of halogen. For example, this could be due to the formation of trap states related to the process of iodine exchange with the gas phase, or to the presence of nucleation centers of iodide rich domains that may be present in higher density for the non-encapsulated sample and may act as recombination centers. It is also important to stress that, in this case, the encapsulated sample undergoes very limited changes in optical absorption during de-mixing and may explain this difference. The negligible change in photo-generation rate is consistent with an almost constant profile of R* during photo de-mixing. At the same time, a similar qualitative trend for the R* evolution as a function of encapsulation conditions is observed for the data discussed for the 2D films, where the encapsulated film undergoes much more limited variation in R* compared with the non-encapsulated sample, despite a comparably large variation in optical absorption.

Remarkably, regardless of encapsulation, the value of R measured for the 3D films shows a (almost) complete reversible behavior after a full photo de-mixing and dark re-mixing cycle. Given the observed degradation of the film without PMMA after photo de-mixing and dark re-mixing, this indicates that the electrical response is not as sensitive as other properties (optical absorption, film crystallinity and phase composition) to changes due to the degradation of the sample. The recovery in R during re-mixing for the 3D films is slower by a factor of about six compared with the de-mixing time. In addition, when plotted on a linear scale, the re-mixing kinetics shows an almost linear trend followed by a plateau when the steady state is reached, possibly suggesting that re-mixing occurs in the form of a zero-order reaction.

In Figure S34 we show temperature-dependent photo de-mixing experiments performed on 3D films with PMMA. The results complement our previous investigation performed on the 2D system. Strikingly, the observation of a decreasing rate of de-mixing for increasing temperature in the 40 – 100 °C range is in contrast with the expected faster kinetics with temperature (as observed for the 2D compound), and it points to a relatively low critical temperature in these 3D systems.

To conclude our investigation, the phase stability of BA-MAPb($I_{0.5}Br_{0.5}$)$_3$ nanocrystalline mixture is investigated using a similar procedure as discussed above. Before and after the in-situ optical absorption measurement during light treatment, the structure of the film is assessed via XRD and UV–Vis. As shown in Figure S36, no change in both the optical and structural properties are observed on illumination of the film for 20 hours with the same illumination conditions used for the study of the 3D and the 2D samples (1.5 mW cm$^{-2}$). This result suggests that photo de-mixing can be suppressed by reducing the grain size, consistent with previous observations[26]. Importantly, these nanocrystalline systems show reduced but still significant ionic conductivity ($\sigma_{ion}$ = 5.2×10$^{-9}$ S/cm and 4.9×10$^{-9}$ S/cm for BA-MAPbI$_3$ and BA-MAPbBr$_3$ at 40 °C) compared with the 3D counterparts ($\sigma_{ion}$ = 8.0×10$^{-9}$ S/cm and 1.1×10$^{-7}$ S/cm for MAPbI$_3$ and MAPbBr$_3$ at 40 °C) (Figure S37-38). Such reduction is due to a larger activation energy of ion transport. Overall, our investigation demonstrates that the lack of de-mixing in mixed-halide nanocrystals is not due to the inability of ionic defects to migrate

### III. Calculations of ionic defect formation and experimental validation

We now consider theoretical aspects that can aid us in elucidating the mechanism and the driving force of the photo de-mixing process in mixed halide perovskites. Specifically, we focus on factors (dimensionality and halide composition) that influence the enthalpies of the systems investigated in this study. We refer to the difference between the enthalpy evaluated for the de-mixed and mixed states as the driving force for de-mixing. One contribution to such driving force concerns the difference in energy of the band edges in the mixed and de-mixed phases, referred to here as the electronic driving force for photo de-mixing. To evaluate additional contributions to the driving force, we consider the role of formation energies associated with ionic defects that may form on interaction with the electronic charge carriers. Such formation energies are, in general, different in the mixed and the de-mixed state and can therefore play a role in the enthalpy change of the system upon photo de-mixing. We refer to such contribution as the ionic driving force in photo de-mixing. To assess the role of the electronic and ionic driving force in photo de-mixing, we perform DFT-based calculations on both 2D halide perovskites (based on PDMA) and 3D halide perovskites (based on MA). The 50% bromide 50% iodide perovskite is used here to discuss the properties of the mixed phase, while the pure iodide perovskites and pure bromide perovskites are considered for the discussion of the I-rich and Br-rich phase, for simplicity.

**Electronic driving force**. First, we consider the electronic driving force ($\Delta E_{eon}$) that arises from the energy gain due to the hole and electron transfer from mixed halide perovskites ($I_{1-x}Br_x$) to the most energetically favorable (separated) phase.[10-12, 35, 36] The end members x = 0 and x = 1 are considered here. The value of $\Delta E_{eon}$ is calculated as the sum of: the energy difference between the conduction band minimum of the de-mixed phase where electrons are most energetically stable and of the mixed phase; the energy difference between the valence band maximum of the mixed phase and of the de-mixed phase where holes are most energetically stable.

Based on the specific position of the electronic energy levels in the compounds investigated here (Figure 5a, b), $\Delta E_{eon}$ reduces to the following difference in energy gap ($E_g$):

$$\Delta E_{eon} = E_g(x=0) - E_g(x=0.5) \quad (2)$$

The results, shown in Figure 5a and b, clearly indicate the tendency for both electrons and holes to be energetically stabilized within iodide-rich phases for both 2D (Figure 5a) and 3D systems (Figure 5b), in line with previous reports on 3D perovskites.[37] This is also consistent with the electronic conductivity data shown in 2D (Figure 5j, $\sigma_{eon}$, yellow circle) and 3D systems (Figure 5k), assuming similar electronic mobilities across the different halide compositions. The electronic driving force for 2D perovskites (due to hole and electron transfer from mixed to pure iodide perovskites, ~ 0.2 eV) is smaller than that of 3D perovskites (~ 0.3 eV).

**Defect formation energy calculations.** To shed light on the role of defects that may arise as a result of the interaction of electronic charge carriers with the ionic sublattice in the photo de-mixing, we perform defect formation energy calculations following the procedure of our previous studies.[16, 17] Several types of point defects including halogen interstitials (H centers, $X_i^\times$), halogen vacancies (F centers, $V_Y^\times$), halide vacancy-interstitial (close Frenkel pairs, ($V_x^\times X_i^\times$)) and $V_k$ centers are considered (Figure 5c). The formation energies of the $V_k$ center involve the dependence on the Fermi level which makes the comparison of different systems more challenging. Therefore, we focus here on neutral defects first and subsequently address the complexity of $V_k$ center calculations. The formation energies of the above-mentioned defects are calculated in different halogen surroundings (Figure 5d, e), including iodine and bromine defects in mixed bromide-iodide perovskites (here referred to as $I_{(x=0.5)}$, $Br_{(x=0.5)}$, respectively), and halogen defects relevant to pure iodide perovskites and bromide perovskites (labelled as $I_{(x=0)}$, $Br_{(x=1)}$, respectively).

First, we discuss the results of these formation energies for the 2D PDMAPb($I_{1-x}Br_x$)$_4$ perovskites, given in Figure 5d and in Table S1–S2, and their implications. The formation energy of an iodine H center in the mixed compound is larger than in the pure iodide compound, which can be related to the smaller size of the inorganic lattice in the mixed compound, as shown by the volume evolution in Table S2. When it comes to Br interstitials, in general, these exhibit lower formation energies than their iodine counterparts, an effect that can be attributed to the smaller size of Br atom compared to I. However, no clear trend in formation energies between the Br interstitial in the mixed and in the pure compound is found, with some formation energies for the mixed compound lower and some similar to the one of the pure compounds. In the case of F centers, the difference between formation energies in mixed and pure compounds for both halogen species is rather negligible. However, the formation energies of Br vacancies are larger than the formation energies of I vacancies, which can be related to a stronger binding of Br to Pb compared to I, consistent with the increased Coulomb interaction with a smaller sized anion and shorter Pb–X (X=I, Br) distances. The formation energies of iodine Frenkel pairs show that such defects are more likely to occur in pure iodide compounds than in the mixture, while the opposite trend is found for bromine defects (following the trend obtained for the H centers).

To compare the formation energies of 3D and 2D systems, we have performed defect formation energy calculations for the MA-based 3D halide perovskites, shown in Figure 5e in Table S3–S4. Results for the H centers show a higher likelihood of a halogen interstitial formation in pure compounds compared to the mixed compound, for both iodine- and bromine-based systems. Comparing the matching pairs of 3D and 2D systems, one can observe lower formation energies for 3D systems. Such an increase in formation energies of iodine interstitials when transitioning from a 3D system to a 2D system is consistent with a previous report.[38] Formation energies of F centers are similar to those in 2D systems, with bromine-based systems having larger formation energies than their iodine-based counterparts. Finally, Frenkel pairs in MAPbI$_3$ and MAPb(I$_{0.5}$Br$_{0.5}$)$_3$ exhibit significantly larger formation energies than in MAPbBr$_3$, which can be related to the change in the symmetry from tetragonal (MAPbI$_3$) to cubic (MAPbBr$_3$), whereby the increase in octahedral tilting (Table S4) leads to more available space for a defect creation in the latter. We additionally note that the interstitial iodine in the mixed compound forms a dumbbell with a Br ion, and not with an I ion, which might influence its formation energy.

In summary, we obtain a consistent trend of iodine H centers having larger formation energies in the mixed compounds compared to the pure iodine analogs for both 2D and 3D systems, which leads to a preferred accumulation of such defects in the pure phase and can thus supply a thermodynamic driving force for photo-induced de-mixing. Moreover, we observe consistently higher formation energies of H centers in 2D systems compared to 3D systems. One should note that the defect calculations for 2D systems are performed only for the sites in the in-plane direction and they do not involve dangling sites. Additionally, the defect calculations of the mixed 3D systems are performed for the energetically more favorable tetragonal phase and not for the cubic phase (these limitations are addressed in the next section).

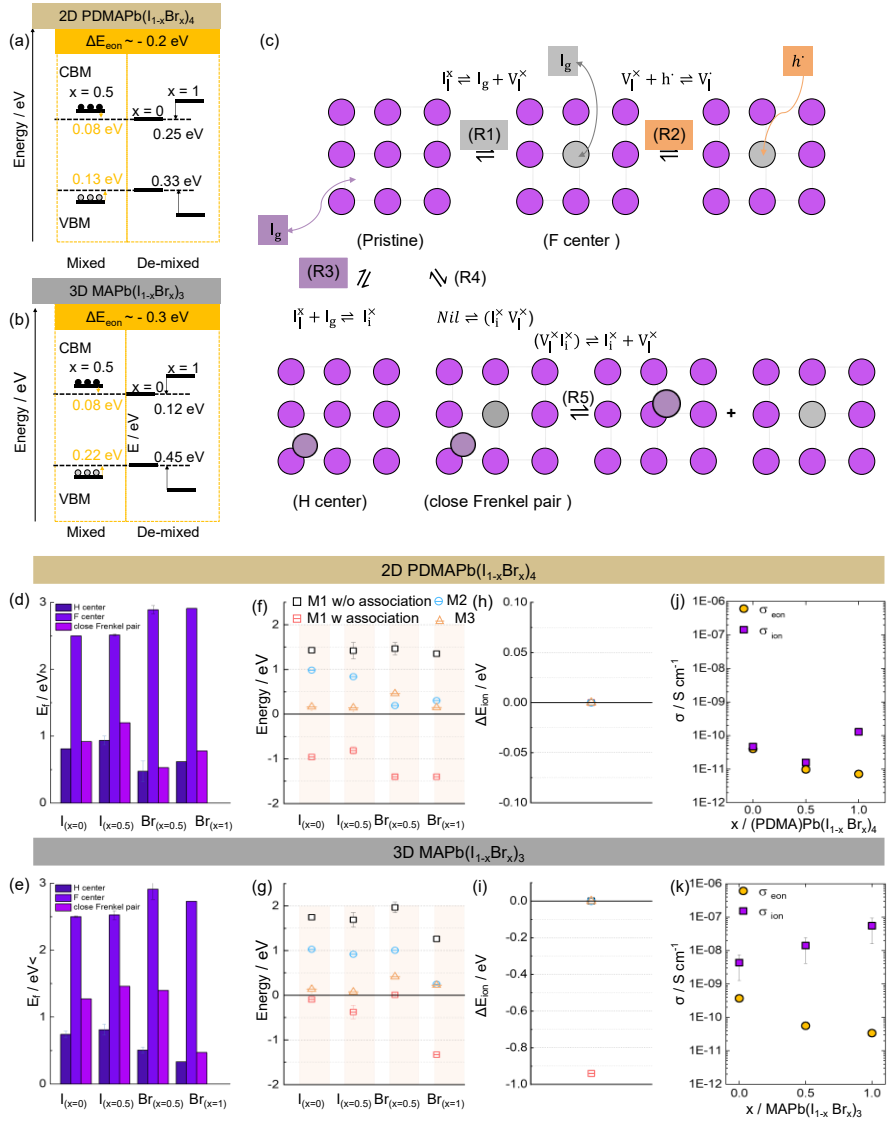

**Figure 5 (a, b) Electronic driving force** of photo de-mixing in (a) 2D PDMAPb($I_{1-x}Br_x$)$_4$ and (b) 3D MAPb($I_{1-x}Br_x$)$_3$. The positions of the VBM and the CBM energy levels are shown with respect to the VBM and the CBM of the pure iodide compounds. The electronic driving force ($\Delta E_{eon}$) for photo de-mixing is calculated from the difference of the electronic energy states between mixed iodide perovskites (x = 0.5) and their end members (x = 0, x = 1), see Eq.2. **(c) Schematic description of the investigated defects and reactions** used for the calculation of the ionic driving force of photo de-mixing (an example of iodine defects in pure iodide perovskites is shown here). (R1) F center formation ($I_I^\times \rightleftharpoons I_g + V_I^\times$, $I_g$ corresponds to an iodine atom in the gas phase). (R2) ionization of a neutral vacancy ($V_I^\times + h^\cdot \rightleftharpoons V_I^\cdot$, the ionization energy of a neutral vacancy minus the bandgap, calculated as the difference between the valence band maximum and the energy level of the vacancy). (R3) iodine incorporation from the gas phase to an interstitial (H center formation). (R4) formation of a close Frenkel pair. (R5) refers to the disassociation of the ($V_I^\times I_i^\times$) Frenkel pair to $V_I^\times$ and $I_i^\times$. The reverse reaction refers to association of $V_I^\times$ and $I_i^\times$ to ($V_I^\times I_i^\times$) Frenkel pair. The schematics are not comprehensive, as only the ionic and electronic defect situation is emphasized. **(d, e) Defect formation energies** of F centers, H centers, close Frenkel pairs in 2D PDMAPb($I_{1-x}Br_x$)$_4$ (d) and 3D MAPb($I_{1-x}Br_x$)$_3$ (e) systems. The defect formation energies are calculated in different halogen surroundings: iodine defects in mixed iodide-bromide perovskites ($I_{(x=0.5)}$); bromine defects in mixed iodide-bromide perovskites ($Br_{(x=0.5)}$); iodine defects in pure iodide perovskites ($I_{(x=0)}$); bromine defects in pure bromide perovskites ($Br_{(x=1)}$). **(f, g) $V_k$ center formation energy** (hole self-trapping on iodide sites leading to the formation of iodine interstitials and vacancies and their further relaxation). Three different approaches are used to calculate such energies in different halogen surroundings. (M1): the $V_k$ center is calculated as a complex defect (consisting of a neutral interstitial and a charged vacancy) and its formation energy is E(R1+R2+R3) (no association) and E(R1+R2+R3-R5) (with association). (M2): the formation energy of the singly-positively charged Frenkel pair is calculated with respect to the pristine structure with an additional hole. (M3): the formation energy of the direct $V_k$ center is calculated with respect to the pristine structure with an additional hole. **(h, i) The ionic driving force** that is related to photo de-mixing is calculated based on comparison of such energy gain in mixed (E($I_{(x=0.5)}$), E($Br_{(x=0.5)}$)), and de-mixed situation(E($I_{x=0}$), E($Br_{x=1}$)). Only negative values are considered, while a driving force of 0 eV is assumed when a positive energy is computed. **(j, k) Electronic conductivity ($\sigma_{eon}$, yellow circle) and ionic conductivity ($\sigma_{ion}$, purple square)** of (j) 2D PDMAPb($I_{1-x}Br_x$)$_4$ and (k) 3D MAPb($I_{1-x}Br_x$)$_3$ perovskite thin films measured in the dark in Ar atmosphere at 60 °C.

**IV. Charged defects calculation and their role in photo de-mixing**

**Ionic driving force.** Based on these data, the energy gain related to the formation of defects upon increase of the electronic charge concentration can be evaluated and compared for the mixed and the de-mixed states. Here, we focus on $V_k$ centers (halogen interstitial associated with a positive vacancy) formation involving either iodide or bromide ions, and we compare formation energies in different situations referenced to the energy of the same system with a free hole. A $V_k$ center formation can occur via hole self-trapping on halide sites further leading to the formation of halogen interstitials and vacancies and their further relaxation (see reactions in Figure 5c, as an example of iodine defects in pure iodide perovskites). This energy is estimated here via three methods:

M1: the $V_k$ center is calculated as a complex defect (consisting of a neutral interstitial and a positively charged halide vacancy) and its formation energy is E(R1+R2+R3) (no association) and E(R1+R2+R3+R5) (with association) (see Figure 5c).

M2: the formation energy of the singly-positively charged close Frenkel pair is calculated with respect to the pristine structure with an additional hole.

M3: the formation energy of the $V_k$ center is calculated directly with respect to the pristine structure with an additional hole.[14, 39] In this case, two neighboring iodine (bromine) atoms are brought closer to each other at the predefined distance, while in the case of M2 method, an iodine (bromine) atom moves to an interstitial position, leaving a vacancy behind (see Figure S1d-e).

The ionic driving force for photo de-mixing deriving from defect formation energies is calculated based on the comparison of the corresponding energy gain in the mixed ($E(I_{(x = 0.5)})$, $E(Br_{(x = 0.5)})$, and de-mixed situation ($E(I_{(x = 0)})$, $E(Br_{(x = 1)})$). The difference between the most energetically favorable $V_k$ center defect in the mixture and in the de-mixed situation is defined as $\Delta E$. Based on this, because only defects with a negative formation energy are expected to form to a significant extent upon increase in electronic charge concentration, we calculate $\Delta E_{ion}$ by considering the difference between the most favorable (and negative) defect formation energies in each phase. If for any of the phases, the energy associated with the most favorable defect is positive, a value of 0 eV is used instead for such phase in the calculation of $\Delta E_{ion}$. If we assume no significant change in the vibrational entropy of the system when considering the defects in the mixed or de-mixed states, $\Delta E_{ion}$ would correspond to a free energy change, without the configurational entropy contribution.

We start with the first method, M1 (Figure 5f, red square with horizontal bisecting line), where the $V_k$ center formation energy is estimated via combining defect reactions, following a similar procedure used to evaluate the $V_k$ center formation energy in $CsPbI_3$.[14, 16] In 2D mixed bromide-iodide perovskites (Figure 5f, M1 with association, and Table S5), we find that $V_k$ center formation involving bromine leads to a more energetically stable state than for the iodine defects ($E(I_{(x = 0.5)})$, $E(Br_{(x = 0.5)})$). When comparing such values to the formation energy of similar $V_k$ centers in the respective pure phases, similar trends are obtained. Considering the energies associated with Br defects (which are more stable in both the mixed and in the de-mixed state compared with their I counterparts), a ~0 eV ionic driving force of photo de-mixing for 2D mixed bromide-iodide perovskites is obtained (Figure 5h upper, red square with horizontal line).

Similar evaluation is also carried out for the 3D case (lower panel Figure 5g, M1 with association and Table S6, red square with horizontal bisecting line). In the mixture, iodine $V_k$ centers are more energetically favorable to form compared with the bromine case ($E(I_{(x = 0.5)}) < E(Br_{(x = 0.5)})$). As for the single halide phases, while iodine $V_k$ centers are less favorable in the pure iodide environment compared with the mixed phase, a very negative formation energy is obtained for the bromine $V_k$ center in the bromide perovskite. Compared with the lowest formation energy calculated for the mixture, this highlights a significant energy gain (-0.94 eV) by forming bromine $V_k$ centers in the bromide perovskite compared with iodine $V_k$ center formation in the mixture (Figure 6e below, red square with horizontal line). While the transfer of holes in the Br-rich phase is unfavorable from an electronic energy level point of view, the stabilization of holes within such defects provides an overall negative free energy change. This analysis suggests that, while electronic effects may dominate the driving force for photo de-mixing in the 2D system investigated here, the same process occurring in the 3D $MAPb(I_{0.5}Br_{0.5})_3$ may be driven by both an electronic and an ionic driving force. It is important to note that the above-discussed $V_k$ center formation energy calculations are based on several assumptions, each of which can influence the results to a certain extent. Firstly, the formation energies of individual H and F centers and Frenkel pairs are dependent on the local environment of a defect within a structure (as shown in Tables S1, S3, S5, and S6). This issue has been partially addressed by performing defect formation energy calculations at different atomic sites (shown as the error bar in Figure 5g). Secondly, the stabilization energies by the Frenkel pair association are calculated for the neutral defects (formation energies of charged Frenkel pairs are given in Table S8 and the corresponding ionic driving forces are shown in Figure 5 as M2) and do not account for a possible charge transfer between the interstitial and the vacancy within the Frenkel pair. This explains the more negative values obtained with method M1 compared with the other methods. Finally, the effects of electron trapping and its contribution to the ionic driving force are excluded from the analysis, which will further decrease the overall free energy. Despite these limitations, our calculations provide a systematic analysis of possible factors influencing the ionic driving force, allowing the comparison between different systems (2D vs 3D and mixed vs pure).

We also calculated the $V_k$ center defect formation energy for $CsPbX_3$ systems (X = I, Br), which pointed to a significant tendency of iodide $V_k$ center to form compared with the bromide counterpart (results for $CsPbI_3$, comparison with Ref[14] and $CsPbBr_3$ are shown in Table S7). These results emphasize that changes in A-cation have important repercussions on such trends, while still stressing the significance of ionic effects in the overall energetics. Furthermore, our results highlight the importance of evaluating the energy associated with the formation of photo-induced defects for both the initial and the final state to obtain an estimate of the ionic driving force. Indeed, in the systems investigated here, $V_k$ center formation appears to be a favorable process also in the mixture, depending on the method used.

Additionally, we complement our study by looking into the formation energies of a direct $V_k$ center which consists of two halogen ions brought close to each other in the presence of a hole (Table S9 and Figure 5f-I as M3). The formation energies indicate that such a process is unfavorable, and it is found to be generally more unfavorable for the bromine-based defects. It is important to note that the reference structure for these calculations, as well as for charged Frenkel pairs, (calculation denoted by M3 and M2, respectively) is the pristine structure with an additional hole, while the reference structure for the calculations denoted by M1 is the pristine structure with no additional charge.

Herein, we comment on some implications related with the assumptions that we have made in calculating defect formation energies so far. First, as previously mentioned, the defect calculations for 2D systems are performed for the sites in the in-plane direction and they do not involve dangling sites. The corresponding formation energies of H and F centers at dangling sites are given in Table S10. One can observe that the formation energy of the iodine H center at a dangling site in the mixed compound is lower compared to the in-plane sites, while the opposite trend is observed for the bromine H centers for both the pure and the mixed compound. On the other hand, all F centers at dangling sites exhibit higher formation energies compared to their in-plane counterparts. Additionally, Frenkel pairs are found to be unstable for dangling sites for some of the compounds and, due to that, we do not consider them in the analysis. The second assumption concerns the defect calculations of the mixed 3D systems that are performed for the assumed tetragonal phase of the mixed compound and not for the cubic phase. The results for the cubic phase of the mixed compound are given in Table S11. The formation energies of H centers in the cubic phase for both iodine and bromine are lower than in the tetragonal phase, which can be related to the larger available spaces for the interstitial atoms in the cubic structure. A similar trend is observed for Frenkel pairs, while F centers do not exhibit any significant difference from the tetragonal phase. The obtained formation energies of $V_k$ centers are lower than for the tetragonal phase (-0.89 eV and -1.00 eV for the iodine and bromine defects for the mixed cubic structure vs -0.36 eV and -0.00 eV for the mixed tetragonal structure). However, these energies are still higher than the formation energy of the $V_k$ center in the pure bromide compound (-1.33 eV). Therefore, the conclusions made for the ionic driving force in the 3D system would remain valid. Overall, the main conclusions remain mostly consistent despite having different ways of treating $V_k$ centers and thus the ionic driving force. Further work is needed to establish a unifying framework for these properties.

Finally, we note that the values of electronic and the ionic driving forces obtained from this discussion cannot be simply added to obtain a total driving force for photo de-mixing. We expect their contribution to be weighed as part of a more complex model that includes all rate equations of electron-hole and ionic defects generation and recombination. For example, $V_k$ centers are likely to affect the recombination rate between electrons and holes in different ways depending on the ion involved and the ionic environment, a factor that will influence the "residence time" of holes within such trapped states and that should be included in the quantification of the ionic driving force. Despite the limitations of the approach adopted here, our discussion sheds light on the relevance of photo-induced ionic defects in the investigation of the energetics associated with the phase stability of halide perovskites under illumination.

## V. Mechanism of photo de-mixing

In this final section, a microscopic picture of the mechanism of photo de-mixing is proposed (Figure 6a) based on the experimental and computational results shown above. We have shown evidence that photo de-mixing in 2D and 3D mixed halide perovskites occur in multiple stages. In the first stage, a composition broadening close to the initial composition ($Br_{0.5}I_{0.5}$ case is discussed here) occurs, a process consistent with spinodal decomposition of the mixture. This is accompanied by significant decrease in the resistance measured in the dark (apart from 2D perovskites without encapsulation), possibly due to an increase in the concentration of mobile ionic defects, to electronic trapping effects, or photo-induced stoichiometry variations. In the second stage, direct transformation of the mixed halide perovskite to the final composition for either the I-rich or the Br-rich phase (or possibly for both of them) occurs via nucleation and growth of these phases (see schematic free energy surface in Figure 6c).

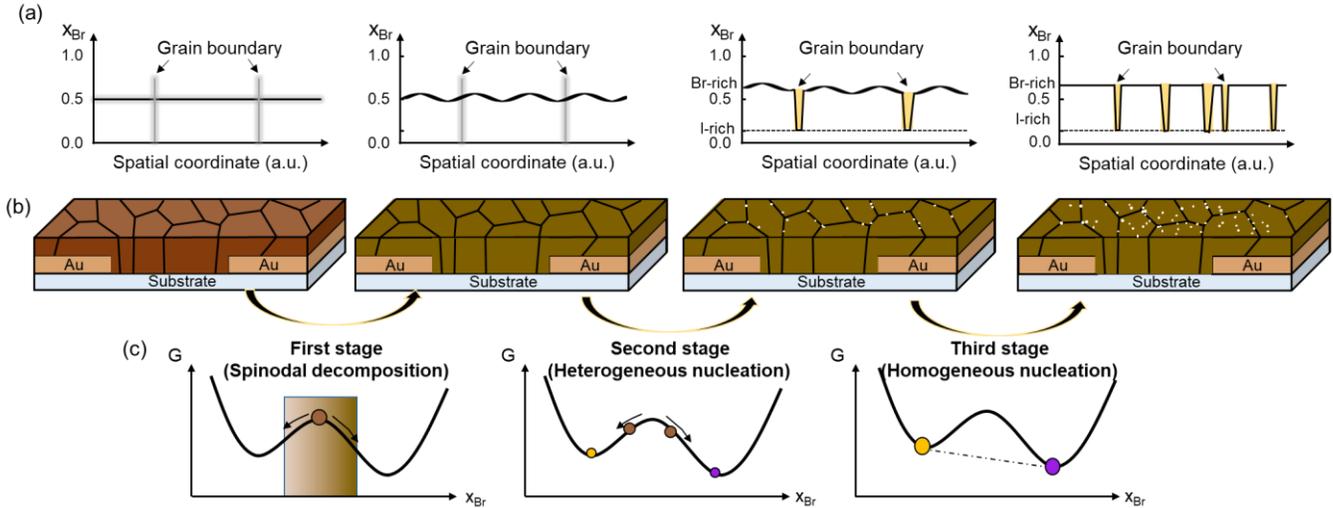

**Figure 6** Microscopic schematic representation of the proposed mechanism for photo de-mixing in the 2D PDMAPb($I_{1-x}Br_x$)$_4$ perovskite, showing spinodal decomposition (first stage) followed by nucleation (second and third stage). Upper panel(a): composition distribution in spatial coordinate; middle panel (b): change in film morphology at different stages; bottom panel(c): compositional variation at different stages illustrated schematically in a free energy (under light) vs composition.

The presence of the nucleation centers along the grain boundaries, but also within the grains, significantly lowers the energy barrier of the final phase formation in the 2D compound, triggering nucleation processes that compete with the spinodal decomposition. Our findings show that the precise details related to the steps involved in such nucleation process and their kinetics (second stage) depend on dimensionality and encapsulation.[11, 14, 15, 36]

For the 2D system, our morphological study points to nucleation of I-rich domains. Formation of both Br-rich and I-rich phases is also observed when illuminating the sample for long time, although under such conditions photo degradation is evident from UV-Vis measurements. For the 3D sample, evidence for nucleation of Br-rich domains and gradual shift in composition in I-rich domains is given by optical spectroscopy measurements and may be related to the predicted enhanced tendency of Br to trap holes and form $V_k$ center defects. We note that, while the transfer of a hole from the mixed phase or an I-rich domain to the Br-rich domain is unfavorable, this trapping makes the process overall thermodynamically favorable, at least for the scenario described in M1 (see previous section). Therefore, besides the well-established electronic contribution ($\Delta E_{eon}$) to the driving force of photo de-mixing, an important role may be played by the photo-induced formation of ionic defects. The stabilizing energy $\Delta E_{ion}$, due to the more favorable formation of such defects in the de-mixed phases compared with the mixed state, represents an ionic driving force to de-mixing, which we show can be larger than the electronic counterpart.

Table 1 summarizes some of the main observations discussed in this work, relating the photo de-mixing dynamics and reversibility to material dimensionality and encapsulation conditions. Interestingly, the trends in optical, structural, and electrical reversibility of the 3D films are rather different to what is observed for the 2D system for the time scale of the measurements considered here. This summary emphasizes that dimensionality as well as encapsulation are key parameters that influence the photo de-mixing behaviour of these mixed-halide systems, and that can be tuned to obtain specific properties in terms of phase-stability and reversibility for applications in optoelectronics and beyond.

**Table 1** Summary of the observed photo de-mixing and dark re-mixing behavior for 2D, 3D and NC perovskite thin films investigated in this study. Reversibility after dark re-mixing is evaluated after 20 h in the dark at the stated temperature.

|  | 2D w/PMMA | 2D w/o PMMA | 3D w PMMA | 3D w/o PMMA | NC |
|---|---|---|---|---|---|
| Occurrence of photo de-mixing | Yes (40 – 100 °C) | Yes (40 – 100 °C) | Yes (40 – 80 °C) | Yes (40 – 80 °C) | No (40 °C) |
| Estimate of critical temperature (1.5 mW cm$^{-2}$, 20 h) | >150 °C |  | ~100 °C |  | <40 °C |
| Number of stages (1.5 mW cm$^{-2}$, 20 h) | 2 | 2 | 1 | 2 | n.a. |
| Optical absorption feature suggesting nucleation of I-rich | Yes | Yes | No | No | n.a. |
| Optical absorption feature suggesting nucleation of Br-rich | No | No | No | Yes | n.a. |
| Structural reversibility | Yes | No | Yes | No | n.a. |
| Optical reversibility | Yes | No | Yes | No | n.a. |
| Electrical reversibility | No | No | Yes | Yes | n.a. |
| Morphological reversibility | - | Partial | - | - | - |

Regarding the effect of encapsulation, exposed surfaces may allow for halogen excorporation and may present enhanced recombination, affecting both the trend in the sample resistance, the morphological evolution and the reversibility of the photo de-mixing process. The systems investigated here (2D, 3D and nanocrystalline thin films) show different grain sizes (Figure 1a). Interestingly, the critical temperature ($T_c$) of photo de-mixing in the three compounds shows a decreasing trend (2D, above 150 °C;[5] 3D: between 40 – 60 °C (Figure S35); Nanocrystals: below 25 °C (Figure S36)) with grain size. This is consistent with the trend in surface energy contribution from the additional phase boundary that would form upon de-mixing to the mixing of free energy. Note that this contribution is expected to be significant only for very small grains, where the energetics is strongly influenced by surfaces and interfaces. Therefore, it might explain only the suppression of the critical temperature associated with the photo-miscibility-gap of nanocrystalline films, but not the variation between 2D and 3D.

**Conclusions**

We investigate the role of electronic and ionic effects in the photo de-mixing of mixed-halide perovskite thin films with different dimensionalities using a set of complementary experimental techniques combined with theoretical calculations. We observe that photo de-mixing occurs in two stages in 2D and 3D perovskite films. An initial stage of de-mixing (<1 hour illumination under the conditions considered here) involves small fluctuations in composition, consistent with spinodal decomposition of the mixed phase, and it is followed by a second stage whereby nucleation events lead to the growth of iodide-rich and/or bromide-rich domains within the film.

We support this picture with ex-situ assessment of the 2D film morphology, where we detect the emergence of de-mixed domains that are rich in iodide forming via heterogeneous nucleation in proximity of the grain boundaries for illumination times > 1 hour. These domains are ~30-70 nm wide and form in the boundary layers of the adjacent grains and close to the film/substrate interface. Homogeneous nucleation of the domains far from the grain boundaries occurs with prolonged illumination. Interestingly, increasing photo de-mixing time results in an increase in the density of such domains while their size remains limited to <100 nm. We have correlated such morphological evolution with in-situ electrical and optical characterization of the thin film response, further confirming the two-stage model. By comparing the response of films with and without encapsulation, we reveal the critical role of the surface in the evolution of the perovskite's (photo-)conductivity and of the phase separation process. We suggest that such difference is due to the modification of the film's surface energy and to the density of the surface defects which influence the rate of electronic charge recombination, domain nucleation rate and solid gas reactions. These factors are expected to affect the mechanism and reversibility of the photo de-mixing process.

A similar investigation carried out for the 3D mixed halide perovskite MAPb(Br$_{0.5}$I$_{0.5}$)$_3$ shows some similarities with the trends observed for the 2D system. The 3D compounds also show an evident 2-stage process in the photo de-mixing. While we cannot confirm this experimentally with microscopy, the second stage seems to involve nucleation of Br-rich phase. Such an observation could be explained based on calculations of the defect formation energies of these systems, whereby an ionic contribution to the driving force for photo de-mixing could derive from the stabilization of holes in bromide $V_k$ centers within Br rich domains, compared to not as favorable formation of such defects in the mixed phase. Discussion of such ionic driving force emphasizes the need to account for the interplay between electronic and ionic charges when evaluating the phase properties of ionic compounds out-of-equilibrium.

For the 3D perovskites, we also present a starker difference in behavior depending on whether the film is encapsulated or not. Specifically, films without encapsulation follow the first and the second stage of photo de-mixing (as described above), the latter being reached already after ~20 minutes at the illumination conditions considered here. For encapsulated films, only the first stage of photo de-mixing occurs, and at a slower rate relative to the non-encapsulated case under the same illumination conditions, with no second stage observed even for illumination time >20 hours. This may be due to the surface passivating properties of PMMA and to hindered halogen excorporation, which could reduce the tendency of the system to nucleate the iodide-rich or bromide- rich domains, although thermodynamic influence of the encapsulant cannot be excluded. Another important observation emphasizing the role of surface energy in the phase stability of these compounds is the fact that no photo de-mixing is detected for the nanocrystalline perovskites, as also shown in previous reports. Here, the energy cost of creating a phase boundary in such a nanosized system is expected to depress the critical temperature to a value below room temperature. Quantification of the electrical properties of the iodide and bromide nanocrystalline perovskite films show significant, although lower than their 3D counterparts, ionic conductivities. This result points towards a thermodynamic origin for the phase stability of nanocrystalline mixed-halide perovskites at temperatures close to ambient.

This study demonstrates a multi-method approach to investigate photo de-mixing experimentally, emphasizing the role of dimensionality and encapsulation on the mechanism, and on the extent and reversibility of such phase-instability. Furthermore, our theoretical analysis provides new insights into the electronic and ionic effects that contribute to the driving force associated with the phase instability of halide perovskites, but also of other ionic systems, when exposed to light.

**Acknowledgement**

Funding Sources

This work was performed within the framework of the Max Planck-EPFL Center for Molecular Nanoscience and Technology. DM is grateful to the Alexander von Humboldt Foundation for funding. EK's research was partly performed in the Center of Excellence of the Institute of Solid-State Physics, University of Latvia, supported through European Union's Horizon 2020 Framework Program H2020-WIDESPREAD-01-2016-2017-TeamingPhase2 under grant agreement No. 739508, project CAMART2, and partly supported by the M ERA NET project HetCat. U.R. gratefully acknowledges funding from the Swiss National Foundation (grant N. 200020_219440) and computational resources from the Swiss National Computing Centre CSCS. M.G. acknowledges financial support from the Günes Perovskite Solar Cell A.S. company, Adana, Turkey and from the Innovation cheque supported by Innosuisse funding application no. 76256.1 INNO-EE.

ACKNOWLEDGMENT

We thank the Nanostructuring Lab (NSL) and the Stuttgart Center for Electron Microscopy (StEM) at the Max Planck Institute for Solid State Research for technical support. Ya-Ru gratefully acknowledges Jürgen Weis and his team (Bernhard Fenk and Ulrike Waizmann) for their technical assistance and support during her doctoral studies. We are grateful to Helga Hoier and Armin Sorg for XRD measurements, and to Florian Kaiser, Rotraut Merkle, Udo Klock, and Kathrin Küster for technical assistance. Special thanks also go to Rotraut Merkle for the internal review of this work and for her constructive comments.

# Relating the dynamics of photo de-mixing in mixed bromide-iodide perovskites to ionic and electronic transport

Ya-Ru Wang,[1,6] Marko Mladenović,[2,4] Eugene Kotomin,[1,5] Kersten Hahn,[1] Jaehyun Lee,[1] Wilfried Sigle,[1] Jovana V. Milić,[3] Peter A. van Aken,[1] Ursula Rothlisberger,[2] Michael Grätzel,[1,3] Davide Moia,[1,*] Joachim Maier[1,*]

[1] Max Planck Institute for Solid State Research, Stuttgart, Germany.

[2] Laboratory of Computational Chemistry and Biochemistry, Institute of Chemical Sciences and Engineering, École Polytechnique Fédérale de Lausanne (EPFL), Lausanne, Switzerland.

[3] Laboratory of Photonics and Interfaces, École Polytechnique Fédérale de Lausanne (EPFL), Lausanne, Switzerland.

[4] Integrated Systems Laboratory, Department of Information Technology and Electrical Engineering, ETH Zurich, Zurich, Switzerland

[5] Institute of solid-state physics, University of Latvia, LV 1063 Riga, Latvia

[6] Technische Universität München, München, Germany.

*moia.davide@gmail.com, office-maier@fkf.mpg.de

*Mixed-halide perovskites, Photo de-mixing, dark re-mixing, ionic and electronic transport, halide compositions, dimensionality, surface encapsulation*

## S1 Experimental

### S1.1 Materials and Synthesis

*Materials:* Lead iodide (PbI$_2$, 99.9985%) and lead bromide (PbBr$_2$, 99.999%) were purchased from Alfa Aesar. Methylammonium iodide (MAI) and Methylammonium bromide (MABr) were synthesized as reported by Im *et al.*[40] Butylamine Hydroiodide (C$_4$H$_{11}$N·HI, 97%) was purchased from TCI.1,4-phenylenedimethanammonium iodide ((PDMA)I$_2$) spacer, 1,4-phenylenedimethanammonium bromide ((PDMA)Br$_2$) spacer were synthesized following the procedure reported for the (PDMA)I$_2$ spacer[41] and the one described below. Dimethyl sulfoxide (DMSO, 99.9%), Dimethylformamide (DMF, 99.8%) and Poly(methyl methacrylate (PMMA, average Mw ~120000) were purchased from Sigma-Aldrich. Chlorobenzene (CB, 99.9%) was purchased in Acros organics. Single crystal sapphire (orientation: (001), C-plane; single polished and double polished) and quartz (molten isotropic quartz) were purchased from CrysTec.

*Synthesis of 2D (PDMA)Pb(I$_{1-x}$Br$_x$)$_4$ precursor solutions:* The precursor solutions with Br content of 0%, 50%, 100% were prepared following the relative stoichiometry of the halides. Specifically, 0.33 M (PDMA)PbI$_4$ or (PDMA)PbBr$_4$ solutions were prepared by dissolving 0.33 mmol (PDMA)I$_2$ ((PDMA)Br$_2$) and 0.33 mmol PbI$_2$ (PbBr$_2$) in the solvent mixture of DMF and DMSO (3:2, (v:v), 200 μL). The precursor solutions for x = 0.5 were prepared by dissolving (PDMA)I$_2$, PbI$_2$, and PbBr$_2$ with stoichiometry of 1: 0: 1.

*Preparation of 2D (PDMA)Pb(I$_{1-x}$Br$_x$)$_4$ thin films:* The film preparation procedure was conducted in Ar-filled glovebox under a controlled atmosphere (O$_2$ and H$_2$O < 0.1 ppm). The (PDMA)Pb(I$_{1-x}$Br$_x$)$_4$ films were deposited on quartz substrate by spin coating the precursor solution at 9 rps and 66 rps for 2s and 48s, respectively. The films were annealed at 150 °C for 10 minutes. A PMMA encapsulation layer was deposited on the perovskite films, unless stated otherwise, by spin coating PMMA solution (2.5 wt%, dissolved in chlorobenzene) on the perovskite surface. The coated sample was dried at 40 °C for 2 hours in the dark in the glovebox.

*Synthesis of 3D MAPb(Br$_x$I$_{1-x}$)$_3$ precursor solutions:* The mixed bromide-iodide perovskites precursors were prepared by dissolving (1-x) mmol MAI and (1-x) mmol PbI$_2$ and x mmol MABr and x mmol PbBr$_2$ (x = 0.0, 0.5, 1.0) in 1 ml DMSO. After that, the as-prepared MAPb(Br$_x$I$_{1-x}$)$_3$ precursor solution was filtered by using a PTFE filter (0.45 μm, Whatman) for the preparation of the thin film.

*Preparation of 3D MAPb(Br$_x$I$_{1-x}$)$_3$ film :* All procedures in this part were conducted in an Ar-filled glovebox with well-controlled atmosphere (O$_2$ and H$_2$O < 0.1 ppm). The MAPb(Br$_x$I$_{1-x}$)$_3$ precursor solution are deposited on the Al$_2$O$_3$ (001) substrate by spin coating the as-synthesized precursor solution at 65 rps and 150 rps for 2s and 180s on Al$_2$O$_3$ substrates. During the spin coating, a 300 μL chlorobenzene drop was dropped on the substrate to induce a quick crystallization. Lastly, the films are annealed at 373 K for 2 minutes. PbI$_2$ and PbBr$_2$ thin films are prepared using the same procedure. A PMMA encapsulation layer was deposited on perovskite films for encapsulation experiments by PMMA solution (2.5 wt%, dissolved in chlorobenzene). The coated sample was dried at 40 °C for 2 hours in the glovebox.

*Synthesis of nanocrystalline BA-MAPb(Br$_{0.5}$I$_{0.5}$)$_3$ precursor solutions:* BA based mixed Iodide and bromide halide nanocrystal perovskites precursor were prepared by dissolving 2 mmol MAI, 0.5 mmol PbI$_2$, 1.5 mmol PbBr$_2$ and 0.4 mmol BAI in 2 ml DMF.

*Synthesis of nanocrystalline BA-MAPb(Br$_{0.5}$I$_{0.5}$)$_3$ thin films:* The film preparation procedure was conducted in an Ar-filled glovebox with a controlled atmosphere (O$_2$ and H$_2$O < 0.1 ppm). The BA-MAPb(Br$_{0.5}$I$_{0.5}$)$_3$ nanocrystal thin films were deposited on sapphire substrate by spin coating the precursor solution at 100 rps for 60 s. A 300 μL chlorobenzene drop was dropped at the time of being spun for 6s to induce a quick crystallization. The films were annealed at 65 °C for 5 minutes.

### S1.2 Experimental apparatus and techniques

**Optical microscope measurements**: The OLYMPUS DSX510 digital microscope in glovebox is used for imaging the morphology of the thin film before illumination, after illumination and after SEM measurements. Dark field imaging mode is used.

**SEM measurements**: The samples are transferred via vacuum shuttle from the argon glove box after optical microscope imaging to a Zeiss Merlin scanning electron microscope, which is used for imaging the change in morphology down to nanometer scale. Four independent electron detectors are used to obtain comprehensive information of the perovskites film: 1) Chamber secondary electron detector (Everhardt-Thornley), 2) In-lens secondary electron (SE) detector (available for < 20 keV), 3) In-lens backscattered electron (BSE) detector (available for < 20keV), 4) 4-segment low-angle scattered electron detector. An accelerating voltage of 1.5 kV was used for the measurement.

**AFM *measurements:*** The AFM measurements are conducted using Bruker's Dimension Icon® Atomic Force Microscope (AFM) System under standard mode in the air.

*Electrical* **measurements:** The in-situ measurements of both the electrical and optical response of the thin films are measured under controlled temperature, atmosphere and light intensity. D.C. galvanostatic polarisation measurements were carried out with a Source Meter (Keithley 6430). A.C. impedance spectroscopy data were acquired with Novocontrol Alpha A with the electrochemical interface.

**Focus Ion Beam (FIB) Processing:** Focused Ga Ion Beam processing was used to prepare thin lamella for TEM investigation. Beam-parameter for preparation the blank (cutting of the lamella from substrate): 30kV10nA. Beam-parameter for preparation the rough lamella: 30kV 2nA. Final polishing with 30kV 200pA/ and 30kV 50 pA.

***TEM-EDX measurements:*** The TEM-EDX measurements are conducted with JEOL ARM 200CF: Scanning transmission electron microscope equipped with a cold field emission electron source, a DCOR probe corrector (CEOS GmbH), a 100 mm² JEOL Centurio EDX Detector, and a Gatan GIF Quantum ERS electron energy-loss spectrometer.

***X-Ray Diffraction (XRD):*** All the XRD patterns were acquired by a PANalytical diffractometer of Empyrean Series 2 (Cu Kα radiation, 40 kV, 40 mA) equipped with a parallel beam mirror and a PIXcel 3D detector. All measurements were conducted in Bragg-Brentano or grazing incidence configuration modules with programmable divergence slits. The Anti scatter slits were used and recorded with a PIXcel 3D detector. The samples that were not encapsulated were mounted in a polycarbonate domed sample holder for protection from ambient atmosphere during the measurements (Panalytical).

## S2 Defect formation energies and ionic driving force
### S2.1 Computational details

DFT calculations are performed using the Quantum Espresso suits of codes.[42] The PBEsol functional[43] is employed for structural relaxation as well as for defect formation energy calculations. Similar trends in defect formation energies are obtained using the PBE functional[44] in conjunction with D3 dispersion correction.[45] Positions of VCM and CBM energy levels are calculated using the PBE0 hybrid functional[46] and by including spin-orbit coupling (SOC) into calculations. Each of the 2D structures, as well as those of MAPbBr$_3$, CsPbI$_3$, and CsPbBr$_3$ are sampled by a 3 x 3 x 3 k-point grid. The pristine and defect-containing structures of MAPbI$_3$ and MAPb(I$_{0.5}$Br$_{0.5}$)$_3$ are sampled by a 4 x 4 x 3 grid. For these two structures, a tetragonal cell containing 4 stochiometric units is used. On the other hand, a 2 x 2 x 2 supercell of a cubic structure is used for MAPbBr$_3$, CsPbI$_3$, and CsPbBr$_3$. Finally, an orthorhombic cell containing 4 stochiometric units is used for all the 2D structures (Figure S1). The kinetic energy cutoffs for the plane wave expansion of the wavefunction and the electron density are 50 Ry and 350 Ry, respectively. The formation energy $E_{form}$ of an H center is calculated as:

$$E_{form} = E_{def} - E_{pris} - \mu_X,$$

where $E_{def}$ and $E_{pris}$ are the energies of a defect-containing and a pristine structure, respectively and $\mu_X$ is the chemical potential of a halogen species, calculated as half of the total energy of an isolated I$_2$ (Br$_2$) molecule. Such a choice for the chemical potential corresponds to the I-rich (Br-rich) case. The formation energy of an F center is calculated as:

$$E_{form} = E_{def} - E_{pris} + \mu_X.$$

Finally, the formation energy of a Frenkel pair is calculated as:

$$E_{form} = E_{def} - E_{pris}.$$

The formation energies of charged Frenkel pairs are calculated in the same way as for the neutral systems, with the difference that in this case, the energies are calculated with respect to the energy of a pristine structure containing a delocalized hole. The same reference energies are used for the direct calculations of V$_k$ centers. To keep V$_k$ centers stable during the calculations, constraints on distances between iodines (bromines) forming the defect are imposed. Distances are kept constant at 3.2 Å (2.85 Å) for the iodine (bromine) defects, and they correspond to the distances between halogen elements in isolated I$_2^-$ (Br$_2^-$) molecules (obtained by calculations).

One should note that the formation energies of both Frenkel pairs as well as disassociated vacancy-interstitial pairs are independent of the choice of the chemical potential. Defect calculations are performed fixing the cell obtained for the respective pristine structures (without any defects). Due to the majority of the investigated defects being neutral and the fact that the formation energies of charged defects are calculated with respect to the corresponding charged pristine structures, we postulate that the interaction between periodic images does not play an important role in defect formation energies.

**S2.2 Defect formation energies**

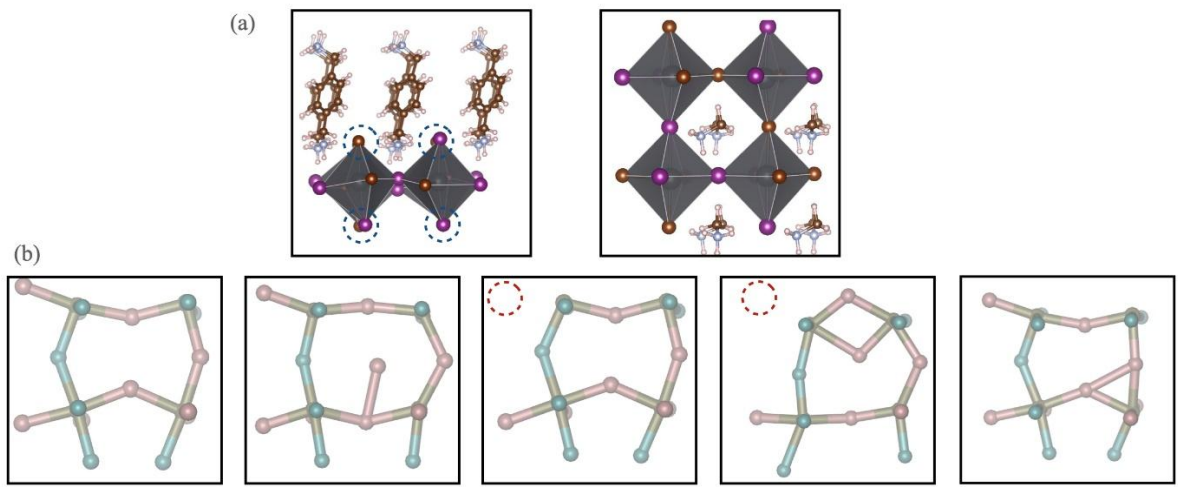

**Figure S1**. (a) Atomic structures of the simulated pristine 2D PDMAPb($I_{0.5}Br_{0.5}$)$_4$ (left) and 3D MAPb($I_{0.5}Br_{0.5}$)$_3$ (right) structures. For 2D PDMAPb($I_{0.5}Br_{0.5}$)$_4$ (for which the defect formation energies are calculated separately), the dangling sites are denoted with blue circles. (b) Illustrations of investigated defects in 2D PDMAPb($I_{0.5}Br_{0.5}$)$_4$, from left to right: pristine structure, H center, F center, close Frenkel pair and $V_k$ center. Position of the halogen vacancies are denoted with a red circle. Iodide atoms are shown in pink, bromine atoms in cyan and lead atoms in olive. PDMA molecules are omitted for clarity.

**Table S1.** Formation energies (in eV) of H centers (halide interstitials), F centers (halide vacancies), and close Frenkel pairs (halide vacancy + halide interstitial) in pure and mixed PDMA-based 2D halide perovskite systems. The values in brackets correspond to defects being placed at different sites.

|  | Iodine defects in PDMAPbI$_4$ | Iodine defects in PDMAPb(I$_{0.5}$Br$_{0.5}$)$_4$ | Bromine defects in PDMAPb(I$_{0.5}$Br$_{0.5}$)$_4$ | Bromine defects in PDMAPbBr$_4$ |
|---|---|---|---|---|
| H center | 0.81 | 1.00 (0.87 0.88) | 0.63 (0.39, 0.32) | 0.62 |
| F center | 2.50 | 2.53 (2.47 2.56) | 2.95 (2.82, 2.96) | 2.91 |
| Close Frenkel pair | 0.92 | 1.20 | 0.53 | 0.78 |

**Table S2.** Cell volume and average octahedral tilting (Pb-I-Pb angles) along two in-plane directions In PDMA-based 2D halide perovskites.

|  | PDMAPbI$_4$ | PDMAPb(I$_{0.5}$Br$_{0.5}$)$_4$ | PDMAPbBr$_4$ |
|---|---|---|---|
| Volume (a.u.) | 12709.48 | 11931.98 | 11038.82 |
| Octahedraltilting (º) | 143.20 | 141.48 | 143.84 |
|  | 144.25 | 144.46 | 143.14 |

**Table S3.** Formation energies (in eV) of H centers (halide interstitials), F centers (halide vacancies), and close Frenkel pairs (halide vacancy + halide interstitial) in pure and mixed MA-based 3D halide perovskite systems. The two values for H centers for MAPbI$_3$ and MAPb(I$_{0.5}$Br$_{0.5}$)$_3$ correspond to defects being placed in the equatorial and the axial plane (shown in brackets), respectively.

|  | Iodine defects in MAPbI$_3$ | Iodine defects in MAPb(I$_{0.5}$Br$_{0.5}$)$_3$ | Bromine defects in MAPb(I$_{0.5}$Br$_{0.5}$)$_3$ | Bromine defects in MAPbBr$_3$ |
|---|---|---|---|---|
| H center | 0.69 (0.79) | 0.73 (0.89) | 0.55 (0.46) | 0.33 |
| F center | 2.49 (2.51) | 2.46 (2.59) | 2.76 (3.06) | 2.73 |
| Close Frenkel pair | 1.40 | 1.27 | 1.46 | 0.47 |

**Table S4.** Cell volume and average octahedral tilting (Pb-I-Pb angles) along three directions in MA-based 3D halide perovskites.

|  | MAPbI$_3$ | MAPb(I$_{0.5}$Br$_{0.5}$)$_3$ | MAPbBr$_3$ |
|---|---|---|---|
| Volume (a.u.) | 13078.50 | 12035.41 | 11112.99 |
| Octahedral tilting (°) | 152.19 | 152.85 | 171.27 |
|  | 152.39 | 151.95 | 162.30 |
|  | 174.70 | 173.42 | 172.07 |

**Table S5.** Formation energies of disassociated vacancy-interstitial Frenkel pairs in PDMA-based 2D systems and their stabilization by association, ionization energies of the neutral halogen vacancies by a hole, and the resulting formation of a $V_k$ center (M1 in main text). The different values for PDMAPb($I_{0.5}Br_{0.5}$)$_4$ correspond to defects placed in different sites of the lattice. All the values are in eV.

|  | Iodine defects in PDMAPbI$_4$ | Iodine defects in PDMAPb($I_{0.5}Br_{0.5}$)$_4$ | Bromine defects in PDMAPb($I_{0.5}Br_{0.5}$)$_4$ | Bromine defects in PDMAPbBr$_4$ |
|---|---|---|---|---|
| H center + F center (no association) | 3.31 | 3.53 (3.34, 3.44) | 3.58 (3.21, 3.28) | 3.53 |
| F center ionization by a hole | -1.88 | -1.93 (-2.10, -1.98) | -1.98 (-1.88, -1.89) | -2.18 |
| H center + ionized F center (no association) | 1.43 | 1.60 (1.24, 1.46) | 1.60 (1.33, 1.39) | 1.35 |
| Stabilization by association | -2.39 | -2.33 (-2.14, -2.24) | -3.05 (-2.68, -2.75) | -2.75 |
| $V_k$ center formation energy | -0.96 | -0.73 (-0.90, -0.78) | -1.45 (-1.35, -1.36) | -1.40 |

**Table S6.** Formation energies (in eV) of disassociated vacancy-interstitial Frenkel pairs in MA-based 3D systems and their stabilization by association, ionization energies of the neutral halogen vacancies by a hole, and the resulting formation of a $V_k$ center (M1 in main text). All the values are in (eV).

|  | Iodine defects in MAPbI$_3$ | Iodine defects in MAPb($I_{0.5}Br_{0.5}$)$_3$ | Bromine defects in MAPb($I_{0.5}Br_{0.5}$)$_3$ | Bromine defects in MAPbBr$_3$ |
|---|---|---|---|---|
| H center + F center (no association) | 3.18 (3.30) | 3.19 (3.48) | 3.31 (3.52) | 3.06 |
| F center ionization by a hole | -1.50 (-1.49) | -1.66 (-1.63) | -1.46 (-1.44) | -1.80 |
| H center + ionized F center (no association) | 1.68 (1.81) | 1.53 (1.85) | 1.85 (2.08) | 1.26 |
| Stabilization by association | -1.78 (-1.90) | -1.92 (-2.21) | -1.85 (-2.06) | -2.59 |
| Vk center formation energy | -0.10 (-0.09) | -0.39 (-0.36) | 0.00 (0.02) | -1.33 |

**Table S7.** Formation energies (in eV) of H centers (halogen interstitials), F centers (halogen vacancies), close and disassociated Frenkel pairs (halogen vacancy + halogen interstitial), ionization energy of the neutral iodine vacancies by a hole, and the resulting formation of a $V_k$ center (M1) in the cubic $CsPbI_3$ and the cubic $CsPbBr_3$.

|  | $CsPbI_3$ | $CsPbBr_3$ | $CsPbI_3$ (Ref [14,23]) |
|---|---|---|---|
| H center | 0.01 | 0.22 | -0.40 |
| F center | 2.39 | 2.94 | 3.00 |
| H center + F center (no association) | 2.40 | 3.16 | 2.60 |
| F center ionization by a hole | -1.12 | -1.15 | -1.60 |
| H center + ionized F center | 1.28 | 2.01 | 1.00 |
| Close Frenkel pair | 0.53 | 0.16 | 0.84 |
| Stabilization by association | -1.87 | -3.00 | -1.76 |
| $V_k$ center formation energy | -0.59 | -0.99 | -0.76 |

**Table S8.** Formation energies of positively charged Frenkel pairs (in eV) in PDMA-based 2D systems and MA-based 3D systems calculated with respect to difference from the charged pristine case. The dependence on the Fermi level is not present in the calculations.

|  | Iodine defects in pure iodine system | Iodine defects in mixed system | Bromine defects in mixed system | Bromine defects in pure bromine system |
|---|---|---|---|---|
| PDMA-2D | 1.14 | 1.01 | 0.43 | 0.53 |
| MA-3D | 1.18 | 1.08 | 1.16 | 0.48 |

**Table S9.** Hole localization energies (in eV) at $V_k$ centers in PDMA-based 2D systems and MA-based 3D systems. Distances between halogens correspond to the distances between halogens in isolated $I_2^-$ ($Br_2^-$) molecules. Values in the brackets correspond to the distance of 3.0 Å, which is the equilibrium distance in $I_2^-$ dumbbell in $PbI_2$. The equilibrium distance in $Br_2^-$ dumbbell in $PbBr_2$ coincides with the equilibrium distance in $Br_2^-$ molecule, therefore, the corresponding formation energies are omitted for the sake of avoiding repetition.

|  | Iodine defects in pure iodide system | Iodine defects in mixed system | Bromine defects in mixed system | Bromine defects in pure bromide system |
|---|---|---|---|---|
| PDMA-2D | 0.33 (0.48) | 0.31 (0.46) | 0.67 | 0.39 |
| MA-3D | 0.30 (0.46) | 0.26 (0.39) | 0.63 | 0.46 |

**Table S10.** Formation energies (in eV) of H centers (halogen interstitials), F centers (halogen vacancies), F center ionization energies and disassociated Frenkel pairs (H center + F center + ionization energy) at dangling sites in pure and mixed PDMA-based 2D halide perovskite systems. Frenkel pairs are found to be unstable at dangling sites.

|  | Iodine defects in PDMAPbI$_4$ | Iodine defects in PDMAPb(I$_{0.5}$Br$_{0.5}$)$_4$ | Bromine defects in PDMAPb(I$_{0.5}$Br$_{0.5}$)$_4$ | Bromine defects in PDMAPbBr$_4$ |
|---|---|---|---|---|
| H center | 0.80 | 0.46 | 0.70 | 0.70 |
| F center | 2.97 | 2.98 | 3.43 | 3.49 |
| H center + F center (no association) | 3.77 | 3.44 | 4.13 | 4.19 |
| F center ionization by a hole | -2.22 | -2.37 | -2.10 | -2.37 |
| H center + ionized F center (no association) | 1.55 | 1.07 | 2.03 | 1.82 |

**Table S11.** Formation energies of H centers (halogen interstitials), F centers (halogen vacancies), close and disassociated Frenkel pairs (halogen vacancy + halogen interstitial) the ionization energies of halogen vacancies by a hole, and the resulting formation of V$_k$ centers (M1 in main text) in the cubic MAPb(I$_{0.5}$Br$_{0.5}$)$_3$. All the values are in eV.

|  | Iodine defects in the mixed system | Bromine defects in the mixed system |
|---|---|---|
| H center | 0.42 | 0.15 |
| F center | 2.41 | 2.64 |
| H center + F center (no association) | 2.83 | 2.79 |
| F center ionization by a hole | -1.74 | -1.61 |
| H center + ionized F center (no association) | 1.09 | 1.18 |
| Close Frenkel pair | 0.85 | 0.61 |
| Stabilization by association | -1.98 | -2.18 |
| V$_k$ center formation energy | -0.89 | -1.00 |

**S3 Structural and optical characterization**

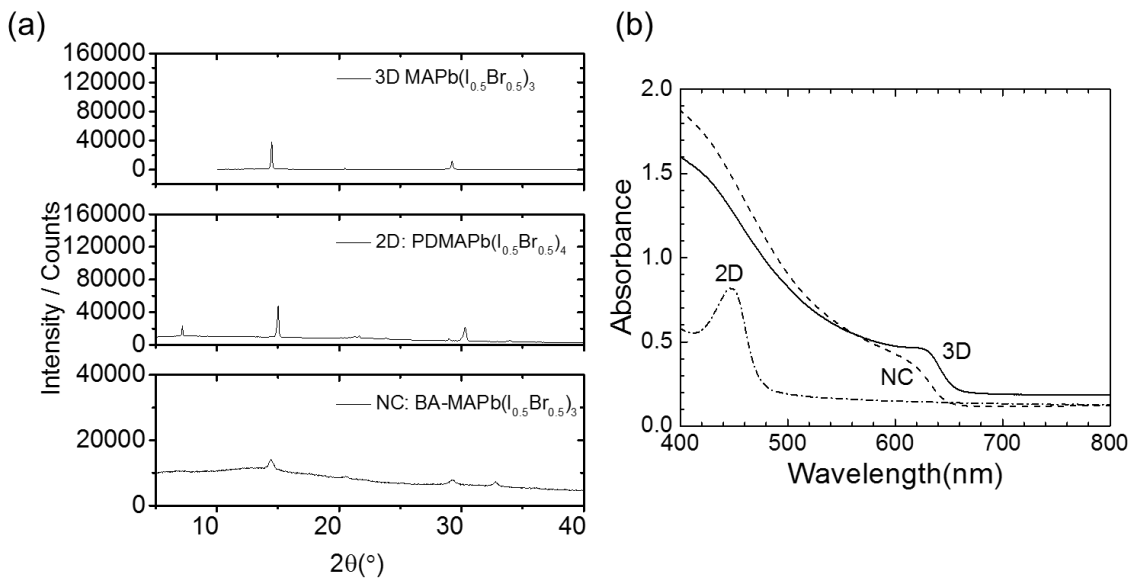

**Figure S2.** (a) XRD (b) UV-Vis spectrum of the mixed halide perovskites with different dimensionalities: three dimensional (3D): $MAPb(I_{0.5}Br_{0.5})_3$ to two dimensional (2D): $(PDMA)Pb(I_{0.5}Br_{0.5})_4$ and nanocrystals (NC): BA- $MAPb(I_{0.5}Br_{0.5})_3$. MA, PDMA and BA represents methyl ammonium ($MA^+$), 1,4-phenylenedimethanammonium ($PDMA^{2+}$) and n-butylammonium ($BA^+$).

## S4 Spectro - electrochemical measurements in mixed bromide – iodide perovskites

### S4.1 Measurements set up and electrode geometry

The in-situ measurements of both the electrical and optical response of the thin films under controlled temperature, atmosphere and light intensity are achieved by using the measurements cells built in-house (Figure S3).

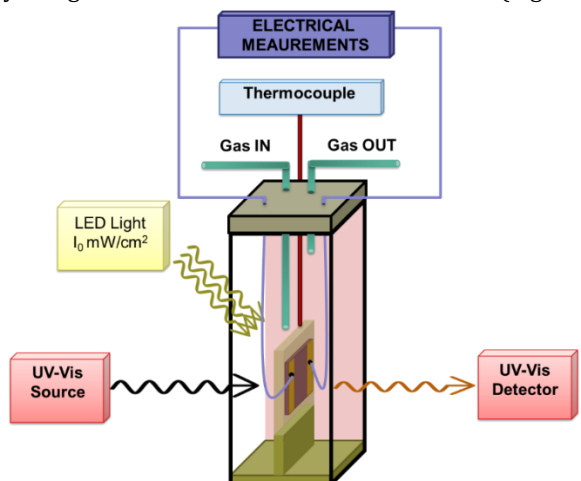

**Figure S3.** Schematic for the measurement cell for UV-Vis & Electrochemical measurements

The Interdigitated electrodes are used for the electrical measurement with the electrode geometry shown in Figure S4. For 2D halide perovskites ((PDMA)Pb($I_{0.5}Br_{0.5}$)$_4$), the electrode geometry of channel length L = 5 µm; Finger width Lc = 5 µm and W overlapping length = 120 cm is used. For 3D halide perovskites (MAPb($_{0.5}$Br$_{0.5}$)$_3$), electrode geometry of L = 10 µm, Lc = 5 µm and W = 80 cm is used.

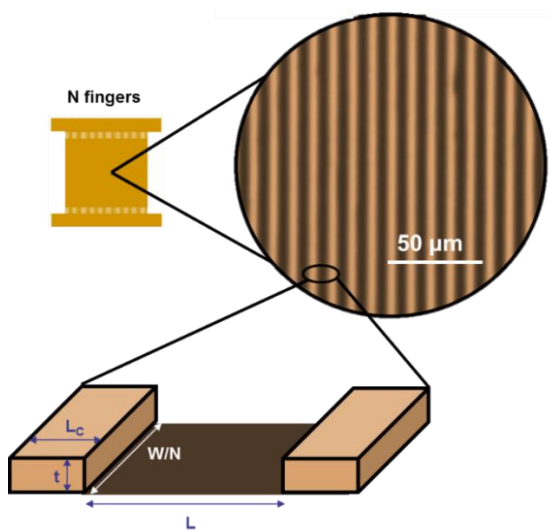

**Figure S4.** Schematic for the interdigitated electrode for electrical measurements. L: channel length; Lc: finger width; t: thickness of the Au fingers; W: overlapping lengths of the fingers; N: number of the fingers.

Simultaneous electrical and optical characterization of thin films during photo de-mixing and dark re-mixing are achieved by careful design of the bias light modulation (Figure S5). Firstly, the modulation of the bias light during photo de-mixing (300 s dark, 900 s light) allows for UV-Vis (in the dark) and impedance (IMP) measurements (conducted every 150 s, both in the dark and under light). In addition, the impedance was taken periodically after switching off the light for ~150 s, a longer time compared with the typical values for photo-generated electron and holes and their recombination. Such design is helpful for making sure that such electrical response is mainly from ionic transport. The UV-vis and IMP measurements are also recorded after switching off the light during the dark re-mixing process.

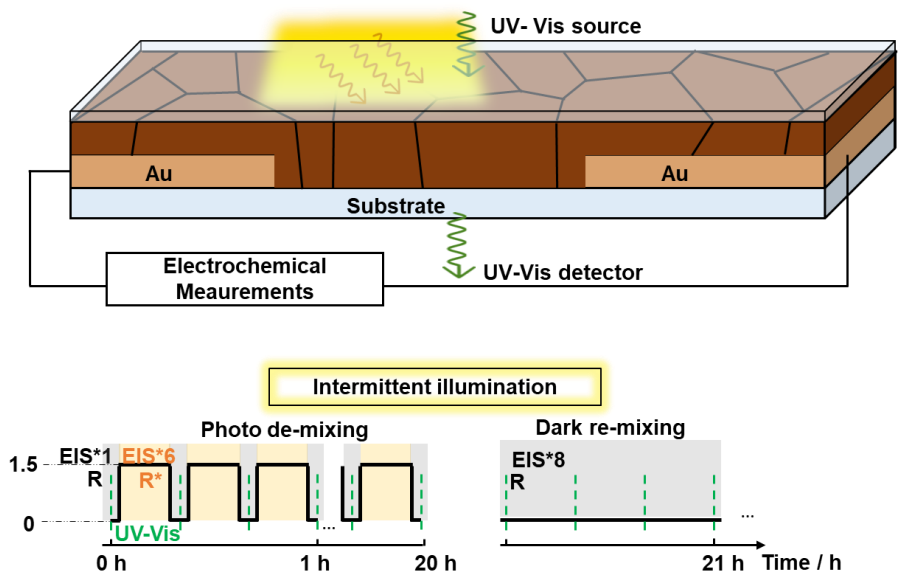

**Figure S5.** Representation of the modulation of the bias light that allows for simultaneous measurements of UV-Vis and impedance during photo de-mixing and dark re-mixing. The light ON/OFF conditions are modulated such that there are three complete cycles per hour (150 s dark, 900 s light, 150 s dark). For each cycle, UV-Vis measurements were taken when the sample is in dark (indicated by black dashed line). Impedance measurements were conducted 2 times in the dark (grey background) and 6 times under light (1.5 mW/cm$^2$, yellow background).

### S4.2 Evaluation of $\sigma_{ion}$ and $\sigma_{eon}$ in the dark

Figure S6 show the DC galvanostatic polarization and impedance measurements of (PDMA)Pb(I$_{0.5}$Br$_{0.5}$)$_4$ thin film in the dark with and without PMMA encapsulation on the surface. From these measurements, ionic and electronic conductivities in the dark can be extracted. In both cases, the ionic conductivity is with almost one order of magnitude higher than the electronic conductivity, suggesting that total resistance (first semi-circle) in the impedance is mainly contributed by the ionic transport. Therefore, the trend of such resistance should be consistent with that of the ionic resistance (R) in Figure 2c.

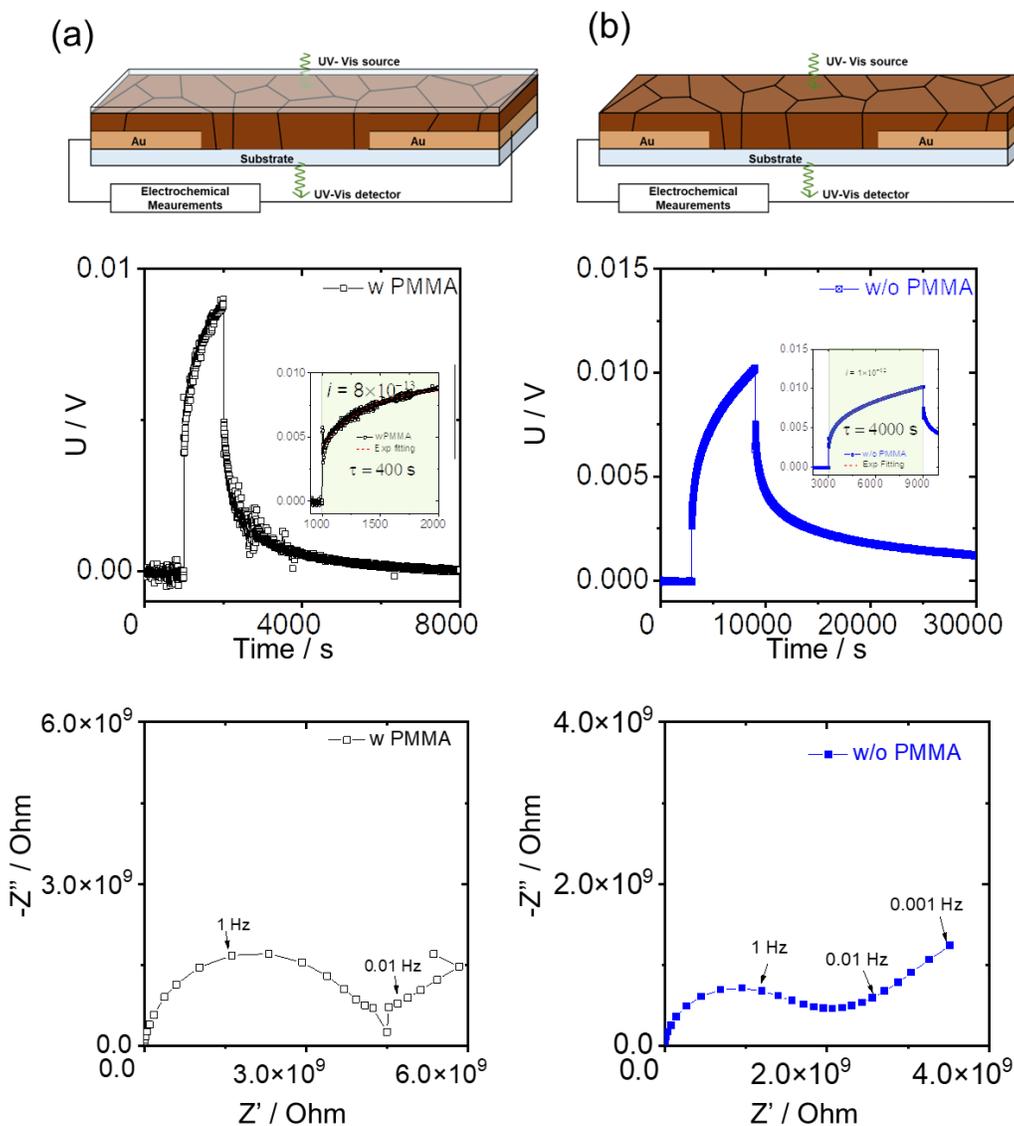

**Figure S6. Electrochemical measurements of** (PDMA)Pb(I$_{0.5}$Br$_{0.5}$)$_4$ thin films in the dark (a) with PMMA and (b) without PMMA encapsulated on the surface.

### S4.3 Evaluation of $\sigma_{ion}$ and $\sigma_{eon}$ under light when quasi-equilibrium is reached

With 20 h illumination, there is limited changes observed from the optical absorption (Figure S7) in both samples with and without PMMA encapsulation on the surface, which we consider that a a quasi-equilibrium situation is reached.

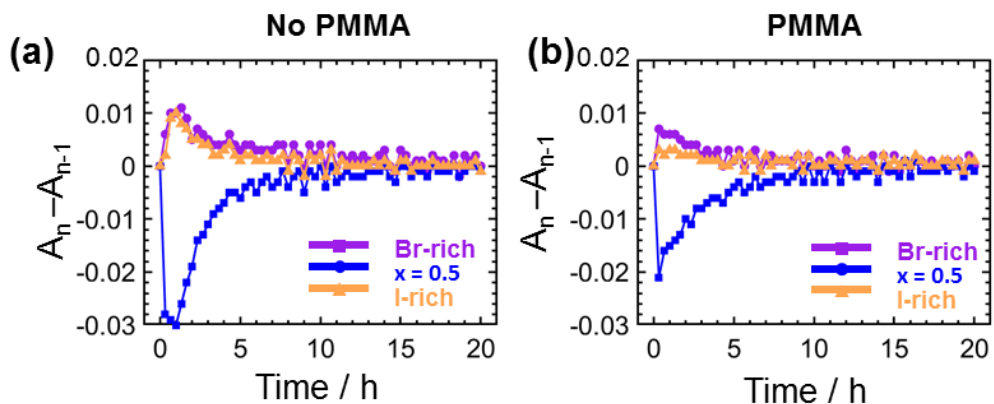

**Figure S7** Change in absorption spectra by subtracting the reference spectrum of the pristine sample from last absorbance spectrum (An-An-1); during photo de-mixing (a) without encapsulation of PMMA, subtracted from absorption data shown in Figure A3.3a. (PDMA)Pb($I_{0.5}Br_{0.5}$)$_4$ (450 nm, blue), Br-rich (406 nm, purple) and I-rich (515nm, yellow) phases.(b) With PMMA encapsulation on surface prior to illumination, subtracted from absorption data shown in Figure 7.12. (PDMA)Pb($I_{0.5}Br_{0.5}$)$_4$ (450 nm, blue), Br-rich (416 nm, purple) and I-rich (517 nm, yellow) The photo de-mixing are conducted under same experimental condition: LED light illumination 1.5 mW/cm$^2$ at 80 °C for ~20 h.

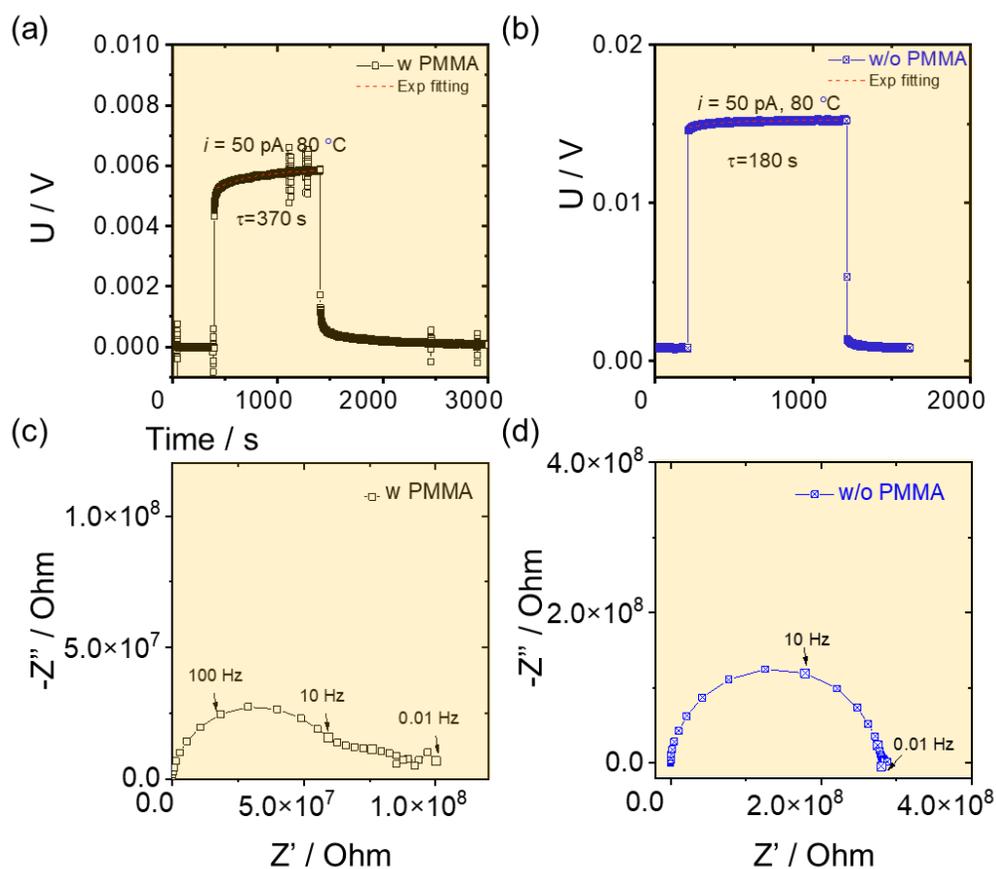

**Figure S8.** Electrochemical measurements of (PDMA)Pb($I_{0.5}Br_{0.5}$)$_4$ thin films under light (after 20 h illumination with 1.5 mW/cm$^2$). (a, c) with PMMA and (b, d) without PMMA encapsulated on the surface. The galvanolstatic polarization and impedance measurements evidenced that the total resistance (first semi-circle) in the impedance is mainly contributed by the electronic transport. Therefore, such resistance can be attributed to electronic resistance (R*) in Figure 2c.

### S4.4 $\sigma_{ion}$ and $\sigma_{eon}$ during photo de-mixing

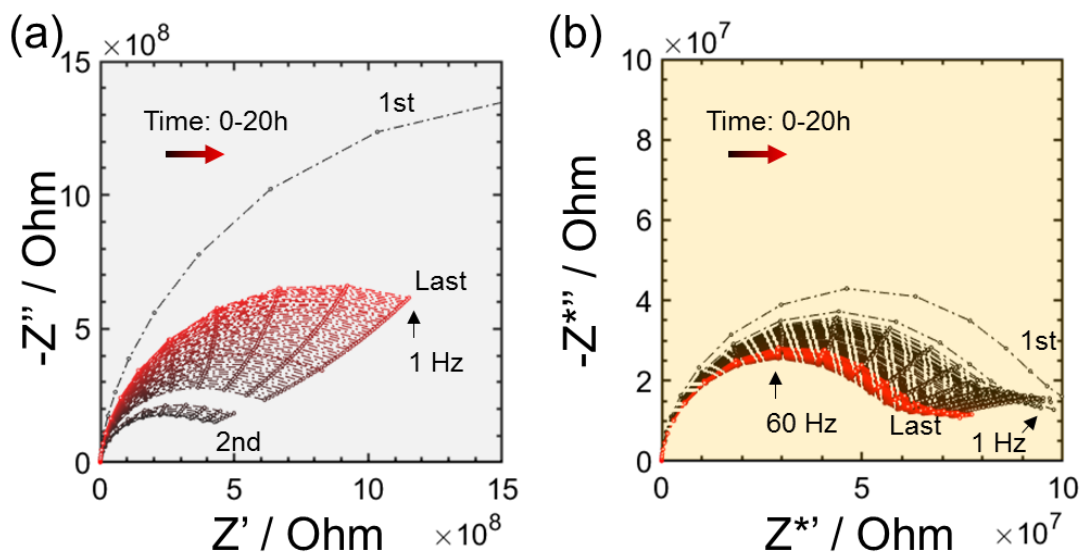

**Figure S9.** Impedance spectra of (PDMA)Pb($I_{0.5}Br_{0.5}$)$_4$ thin film (with PMMA encapsulation, ~ 60 nm): (a) measured in the dark (b) measured under light (1.5 mW/cm$^2$) during photo de-mixing (at 80 °C for 20 h, in Ar atmosphere). The evolution of the resistance extracted from the high frequency semi-circle of the impedance measured in the dark (R) and under light (R*) is shown in Figure 2c.

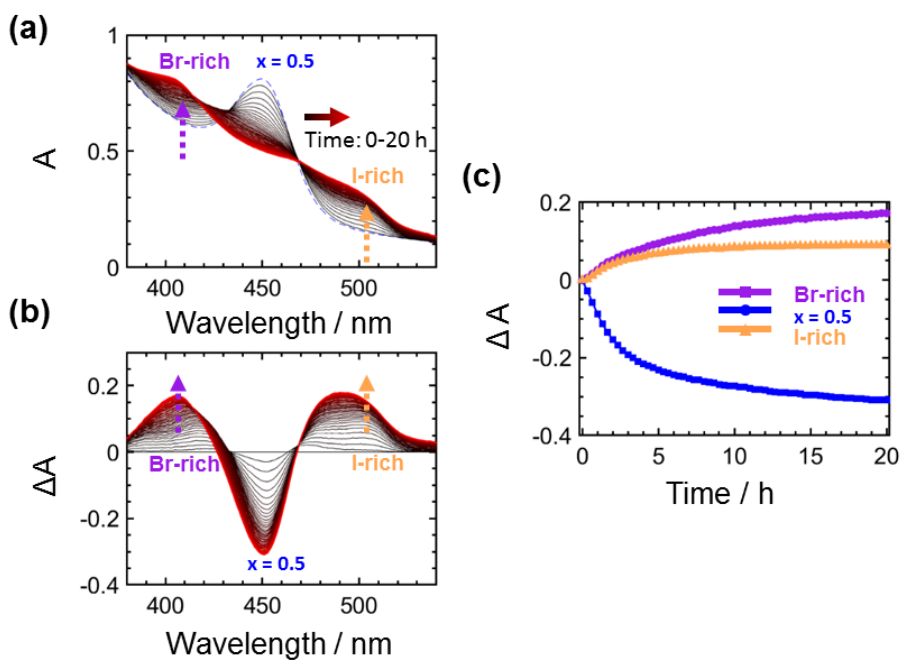

**Figure S10** UV-Vis spectra evolution of (PDMA)Pb($I_{0.5}Br_{0.5}$)$_4$ thin films (without PMMA encapsulation and in Ar atmosphere) during photo de-mixing under light (1.5 mW/cm$^2$) at 80 °C for 20 h (a) Absorbance and (b) Change in absorbance obtained by subtracting the reference spectrum of the pristine sample from each absorbance spectrum shown in (a); (c) Kinetics of the photo de-mixing highlighting the change in absorbance at wavelengths associated to (PDMA)Pb($I_{0.5}Br_{0.5}$)$_4$ (450 nm, blue), Br-rich (406 nm, purple) and I-rich (515 nm, yellow) phases.

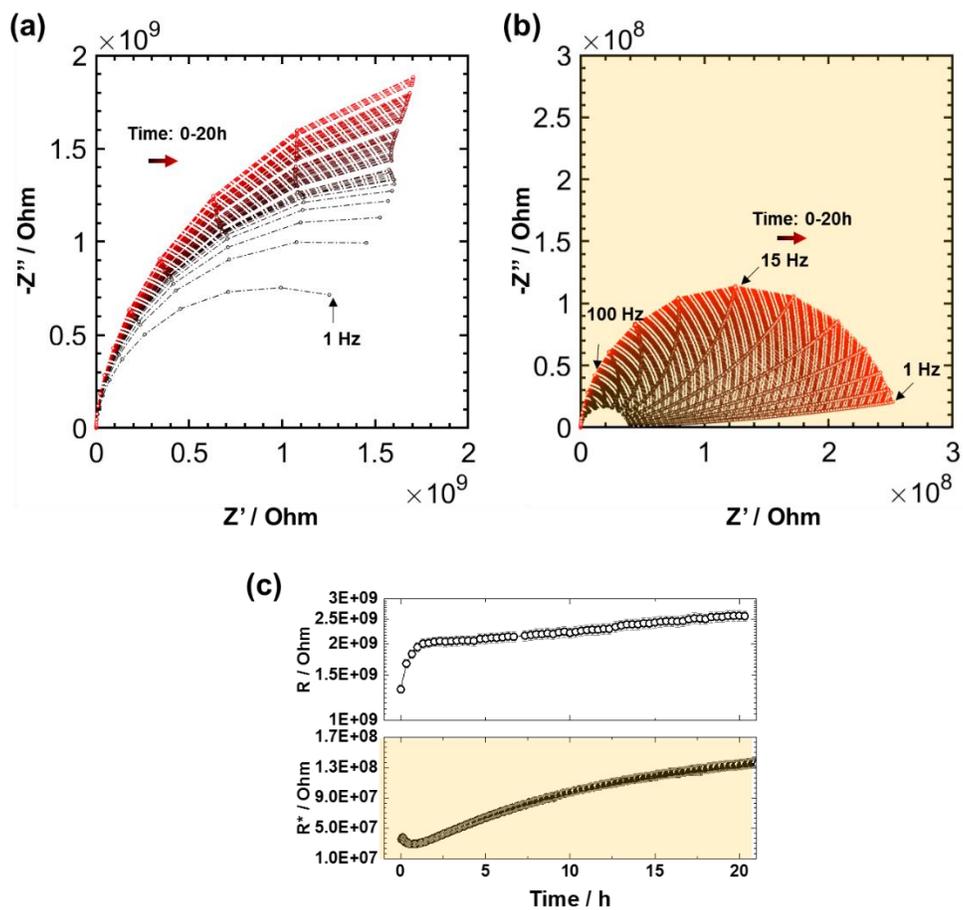

**Figure S11** Impedance spectra of (PDMA)Pb(I$_{0.5}$Br$_{0.5}$)$_4$ thin film (without PMMA encapsulation, in Ar atmosphere) during photo de-mixing (at 80 °C for 20 h) (a) measured in the dark (b) measured under light (1.5 mW/cm$^2$); The measurement frequency range is from 1 MHz to 1 Hz. (c) Change in resistance in dark (R) and under light (R*).

**S5 Microscopic measurements**

**S5.1 Effect of illumination and E-beam exposure – single sample measurments**

To access information on the changes in morphology and phase properties of the perovskite surface pre- and post-illumination, same area of one PDMAPb($I_{0.5}Br_{0.5}$)$_4$ thin film are investigated after different treatment (Figure S12-14). SEM images were taken firstly at pristine state: (a), (b), (c); After the SEM investigation, the film was illuminated under 1.5 mW/cm$^2$ at 80°C for 20 hours, after which another set of SEM images were taken shown in (d), (e), (f). Then the perovskites thin film was taken to glovebox annealed in Ar at 80 °C for 30 days, after which SEM image of (g), (h), (i) were taken in the same area. (a), (d), (g) InLens; (b), (e), (h) SE2; (c), (f), (i) ESB detectors. Two different areas were investigated: area where perovskite thin films are on the Au finger (purple) and area where perovskite thin films are grown on top of Au pad (marked in orange).

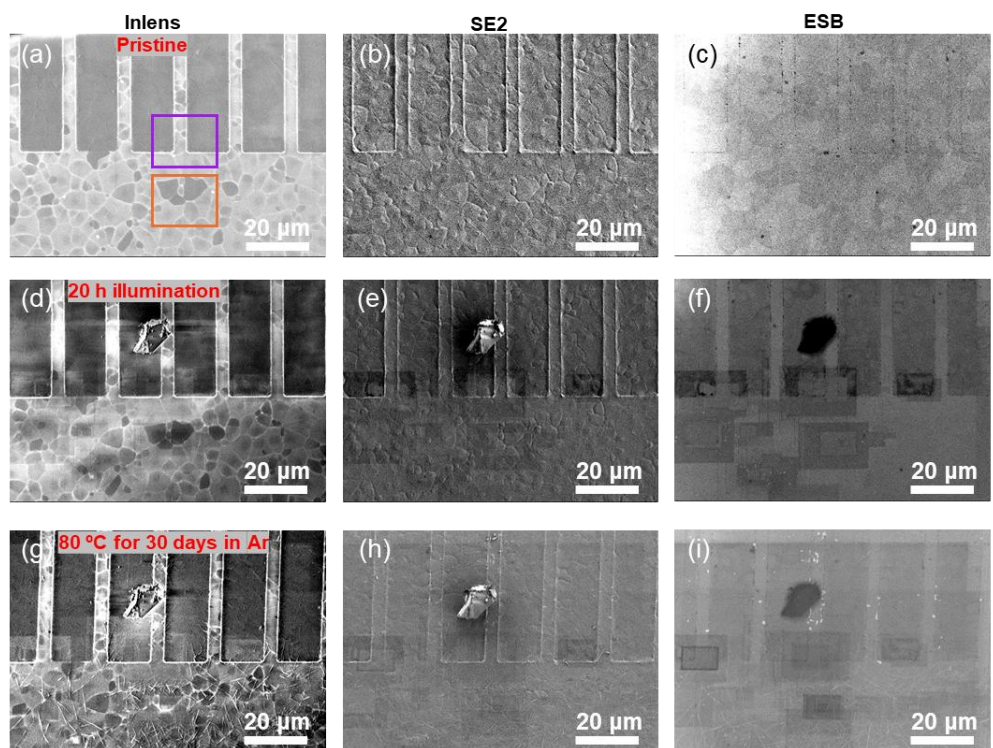

**Figure S12**. The SEM images for a PDMAPb($I_{0.5}Br_{0.5}$)$_4$ film on interdigitated gold electrodes under (a), (b), (c) pristine state; After the SEM investigation, the film was illuminated under 1.5 mW/cm$^2$ at 80°C for 20 hours, after which another set of SEM images were taken shown in (d), (e), (f). Then the perovskites thin film was taken to glovebox annealed in Ar at 80 °C for 30 days, after which SEM image of (g), (h), (i) were taken in the same area. (a), (d), (g) InLens; (b), (e), (h) SE2; (c), (f), (i) ESB detectors. The film is not encapsulated. Magnification: 1K. Note that the square area being marked in purple and orange in Figure S12a will be used to image under higher magnifications in Figure S13 and Figure S14.

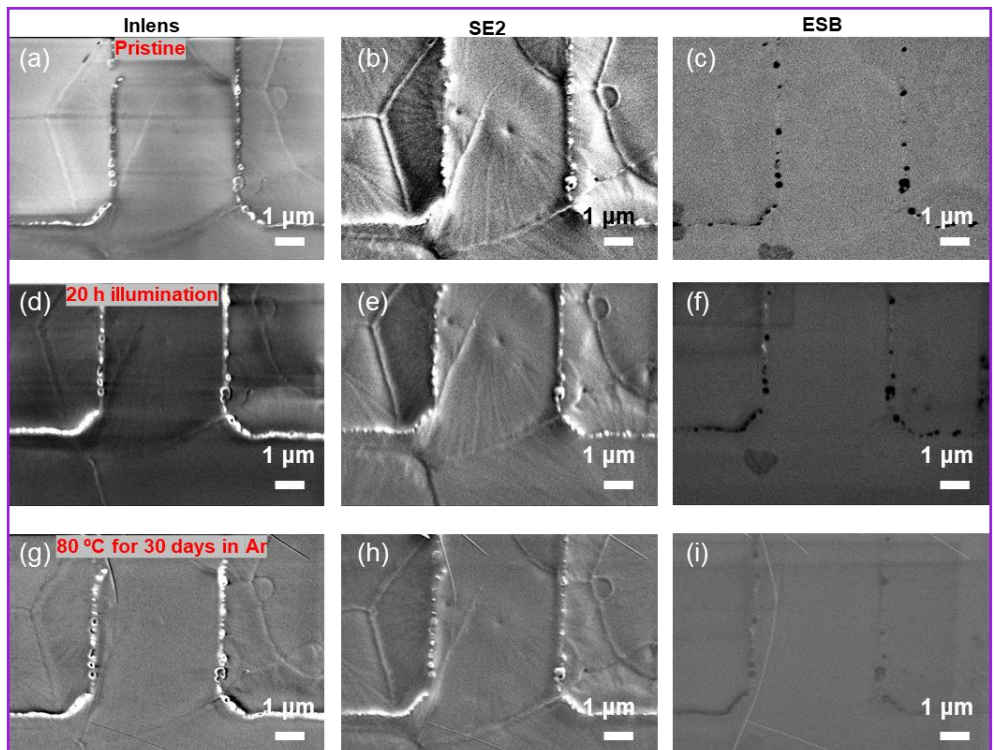

**Figure S13**. The SEM images for a PDMAPb(I$_{0.5}$Br$_{0.5}$)$_4$ film on interdigitated gold electrodes of the purple square marked in Figure S12a: (a), (b), (c) pristine state; After the SEM investigation, the film was illuminated at 1.5 mW/cm$^2$ at 80 °C for 20 hours, after which another set of SEM images were taken shown in (d), (e), (f). Then the perovskites thin film was taken to glovebox annealed in Ar at 80 °C for 30 days, after which SEM image of (g), (h), (i) were taken in the same area. (a), (d), (g) InLens; (b),(e), (h) SE2; (c),(f), (i) ESB detectors. The film is not encapsulated. Magnification: 10K.

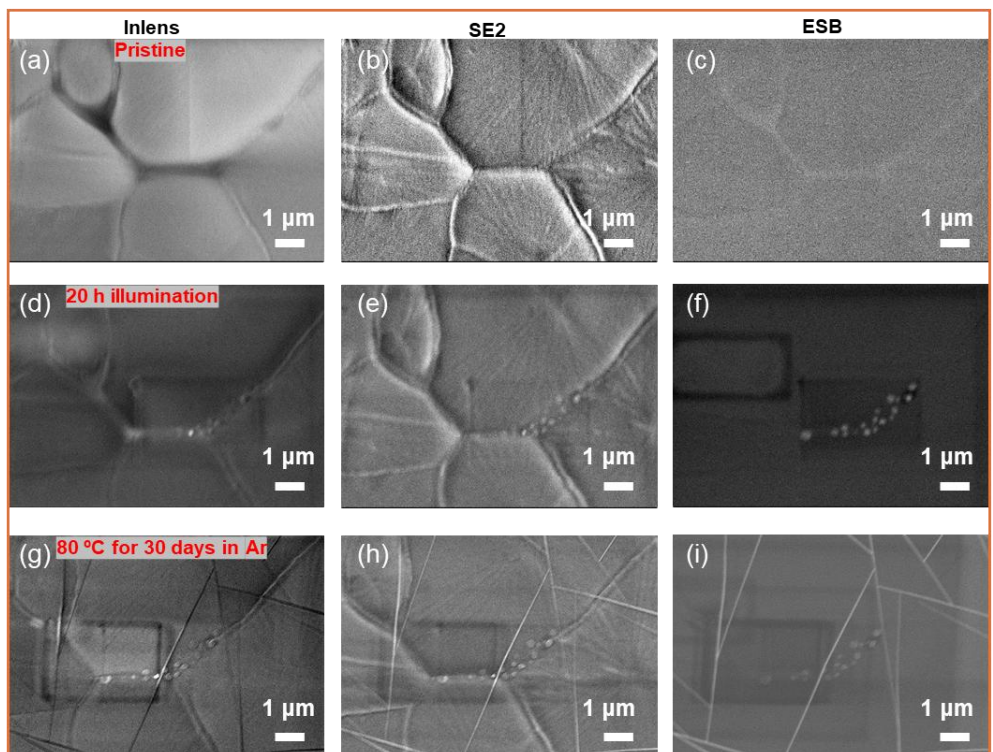

**Figure S14**. The SEM images for a PDMAPb(I$_{0.5}$Br$_{0.5}$)$_4$ film on the interdigitated gold electrodes with Au contact below (marked in orange in Figure S14a: (a), (b), (c) pristine state; After the SEM investigation, the film was illuminated at 1.5 mW/cm$^2$ at 80 °C for 20 hours, after which another set of SEM images were taken shown in (d), (e), (f). Then the perovskites thin film was taken to glovebox annealed in Ar at 80 °C for 30 days, after which SEM image of (g), (h), (i) were taken in the same area. (a), (d), (g) InLens; (b),(e), (h) SE2; (c),(f), (i) ESB detectors. The film is not encapsulated. Magnification: 10K.

Firstly, the perovskite thin films grown on the Au interdigitated electrodes have the similar grain structure of 2D PDMAPb($I_{0.5}Br_{0.5}$)$_4$ thin film as shown in Figure 1(middle), Figure 1e, which assures for the fair comparison with other optical absorption and electrical measurements data. For the perovskite thin film grown on top of Au pad (S13, marked in purple in Figure S12), in general the grains show good continuity with small holes close to the Au interdigitated electrode and perovskite surface due to the 200 nm height difference. After 20 h illumination, only a small bright domain formed close to the holes mentioned above (Figure S13f). Due to the heterogeneous charge distribution between the perovskites and gold electrode, the changes on the grain boundaries can be hindered. The area where perovskite thin films are grown on top of Au pad (Figure S14, marked in orange in Figure S12) however provides such information better due to the more homogeneous distribution of electronic charges on the surface. Small nanodomains are observed close to the grain boundary area shown in Figure S14f after 20 h illumination. To investigate the reversibility of such changes, the same perovskites thin film was taken to glovebox annealed in Ar at 80 °C for 30 days, after which another set of SEM image were taken (Figure S12-14, bottom panel). Surprisingly, both regions show formation of needle-like structure on the regions that was exposed to E-beam due to SEM imaging. Interestingly, the density of these needle like structure increases at the regions where higher dose of E-beam was exposed (Figure S15a). At the region where no E-beam was exposed, such needle structures was not observed at all (See Figure S16, Figure S12), pointing to the effect of E-beam on changing defects in the material, therefore the photo effect and thermal effect. This observation deserves further investigation.

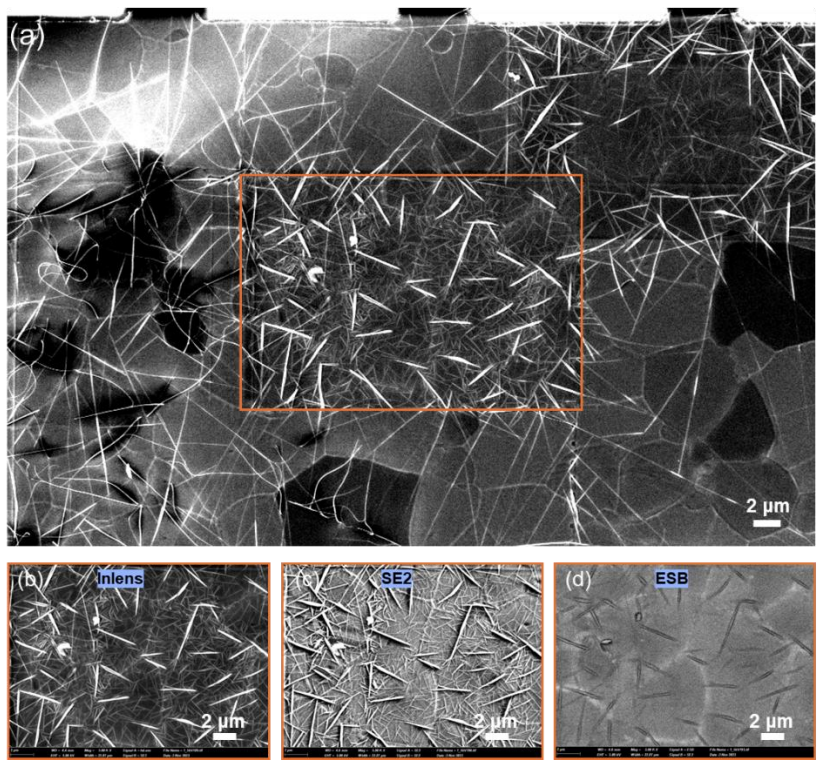

**Figure S15**. SEM image of a PDMAPb($I_{0.5}Br_{0.5}$)$_4$ film after being exposed to light at 1.5 mW/cm$^2$ at 80°C for 20 hours, being annealed in the glovebox at 80 °C for 30 days, after 3 times SEM imaging during which it was being exposed to E-beam (a region shown in Figure S12g). (a) Mag = 1.9 KX; (b-d) Magnified SEM (b) InLens; (c) SE2; (d)ESB of Figure S15a marked with orange square, where higher dose of E-beam was exposed. Mag = 5KX.

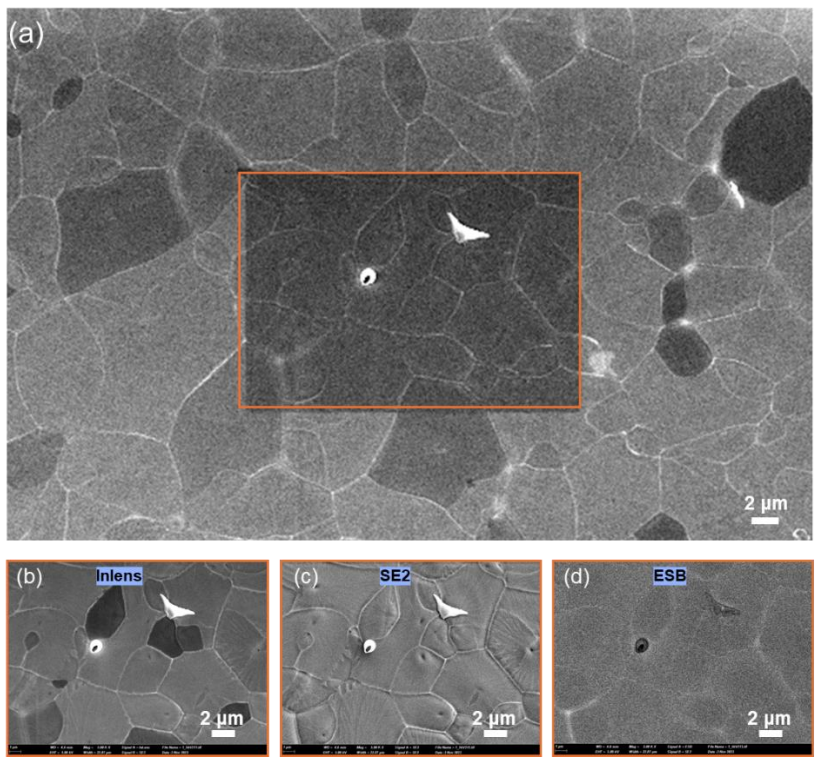

**Figure S16**. SEM image of a PDMAPb($I_{0.5}Br_{0.5}$)$_4$ film after being exposed to light at 1.5 mW/cm$^2$ at 80 °C for 20 hours, being annealed in the glovebox at 80 °C for 30 days, no exposure to E-beam before. (a) Mag = 1.9 KX; (b-d) Magnified SEM (b) InLens; (c) SE2; (d)ESB of Figure S16a marked with orange square, where higher dose of E-beam was exposed. Mag = 5KX.

To minimize the influence of E-beam, ex-situ SEM measurements for these 2D thin films without encapsulation are carried out. Silicon substrates with a thin top layer of $Si_3N_4$ and with small windows where the silicon is removed are used (Figure S16). This choice allows one to perform SEM measurements with good resolution without conductive coating and it also offers the possibility of conducting imaging measurements in transmission mode. To study the changes in perovskite thin film morphology during photo de-mixing, the illumination time was varied (0 h, 1 h, 20 h, 50 h) as shown in Figure 3a (a back scattered electron (BSE) detector is used, see Figures S20–S24 for measurements with other detectors).

Ex-situ SEM measurements are conducted to examine the changes in morphology and phase properties pre- and post-illumination. In this study, silicon substrates that had a thin layer of $Si_3N_4$ on top are used. Each substrate has several "windows" where no silicon is present and consisted only of the $Si_3N_4$ film (15 nm thickness) (Figure S17). The use of $Si_3N_4$ as a substrate provided two advantages: (1) the thin substrate layer allows imaging measurements to be conducted in transmission mode; (2) the window positions (100 × 100 µm) could be used to identify the grain under consideration (Figure S19).

**S5.2 Effect of illumination – multi sample measurements**

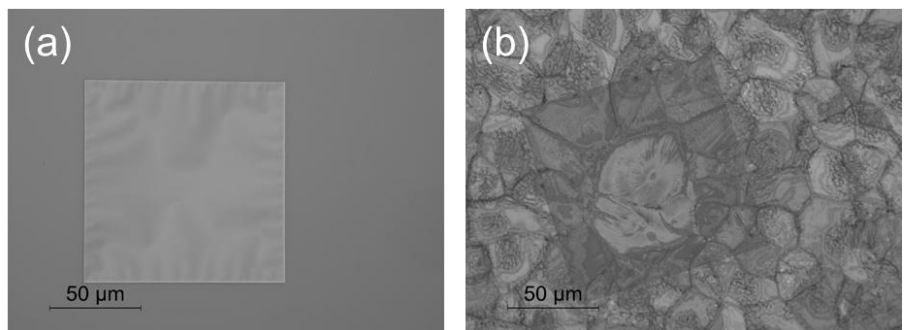

**Figure S17** Optical microscope image of the (a) bare $Si_3N_4$ membrane substrates (b) PDMAPb($I_{0.5}Br_{0.5}$)$_4$ thin films deposited on top of the membrane substrates shown in (a).

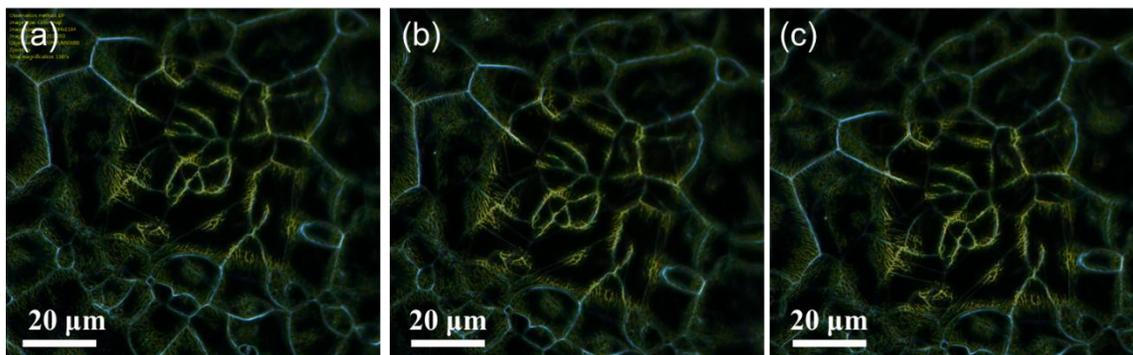

**Figure S18.** Optical image (dark field) for PDMAPb(I$_{0.5}$Br$_{0.5}$)$_4$ thin film (a) pristine; (b) after illumination at 1.5 mW/cm$^2$ at 80°C for 20 hours. (c) After SEM measurements (after approximately 4 -5 hours of exposure to the electron beam).

The morphology of the perovskite thin film prior to illumination on the window area and non-window area are shown in Figure S19. Compared with the non-window area, portions of the film on the window area show more cracks. This can be due to the more favourable release of the strain (generated from different thermal expansion coefficient between the perovskites thin film and substrate during preparation process) to bulk Si compared with thin film Si$_3$N$_4$.

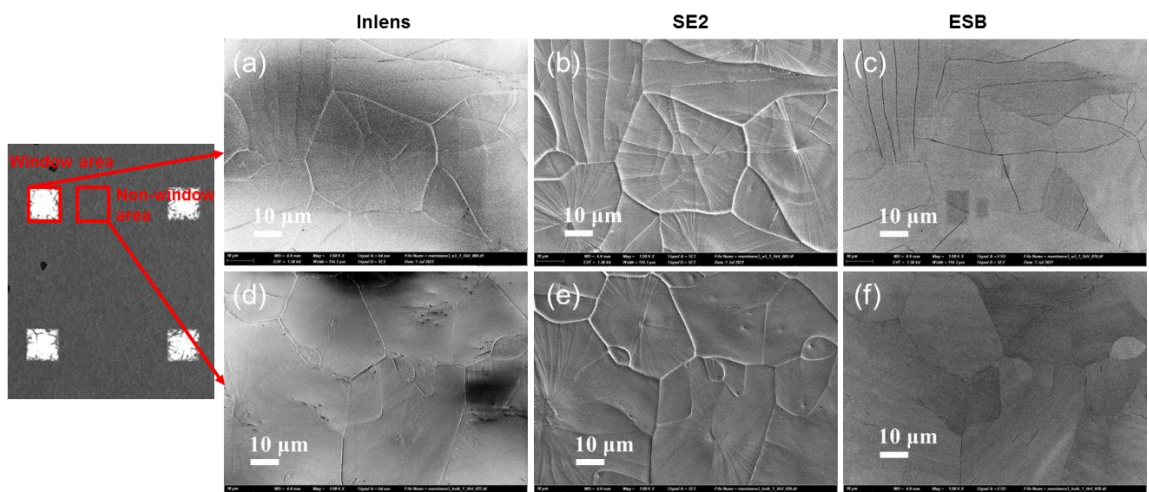

**Figure S19.** SEM of pristine PDMAPb(I$_{0.5}$Br$_{0.5}$)$_4$ thin film (a-c) window area; (d-f) Non window area. (a), (d) InLens; (b), (e)SE2; (c), (f)BSE detectors. Magnification: 1K.

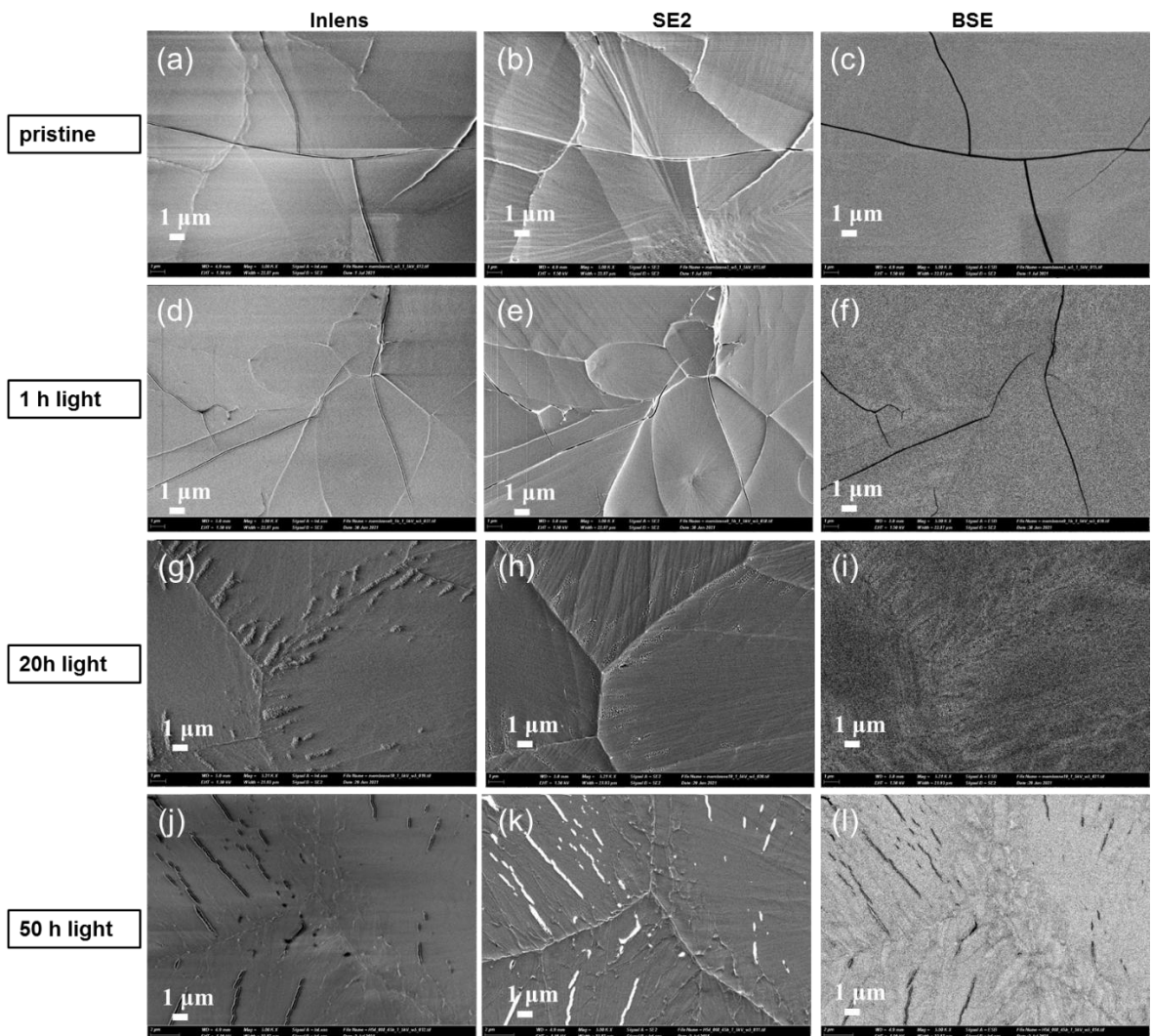

**Figure S20** SEM of PDMAPb($I_{0.5}Br_{0.5}$)$_4$ thin film with different de-mixing time (illumination intensity 1.5 mW/cm$^2$ at 80°C). (a-c) pristine; (d-f) de-mixed for 1 h; (g-i) de-mixied for 20 h; (j-l) de-mixed for 50 h. (a, d, g, j) InLens; (b, e, h, k) SE2; (c, f, i, l) BSE detectors. The imaging was conducted in window area. Magnification: 5K.

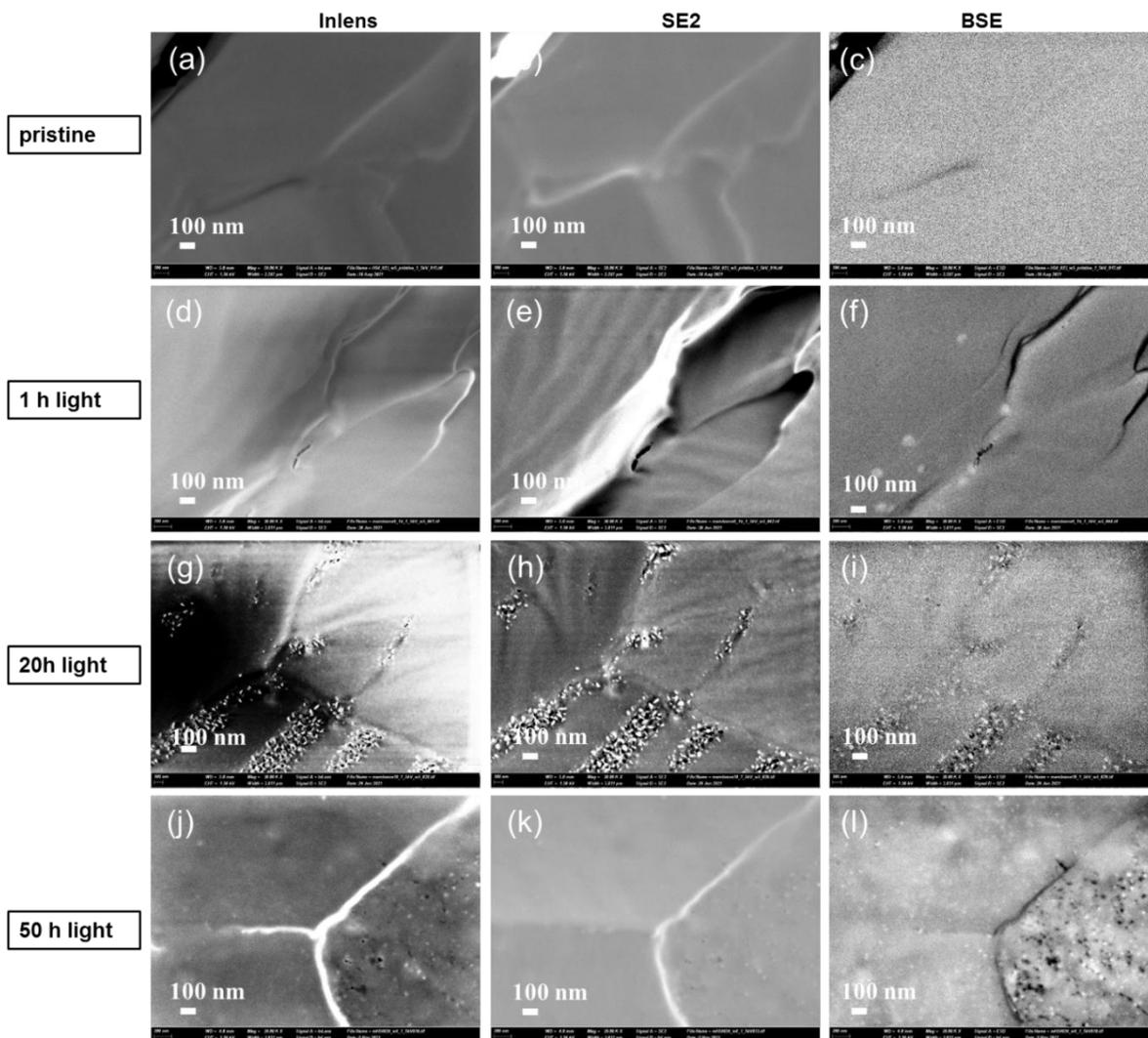

**Figure S21** SEM of PDMAPb(I$_{0.5}$Br$_{0.5}$)$_4$ thin film with different de-mixing time (illumination intensity 1.5 mW/cm$^2$ at 80°C). (a-c) pristine; (d-f) de-mixed for 1 h; (g-i) de-mixied for 20 h; (j-l) de-mixed for 50 h. (a, d, g, j) InLens; (b, e, h, k) SE2; (c, f, i, l) BSE detectors. The imaging was conducted in window area. Magnification: 30K.

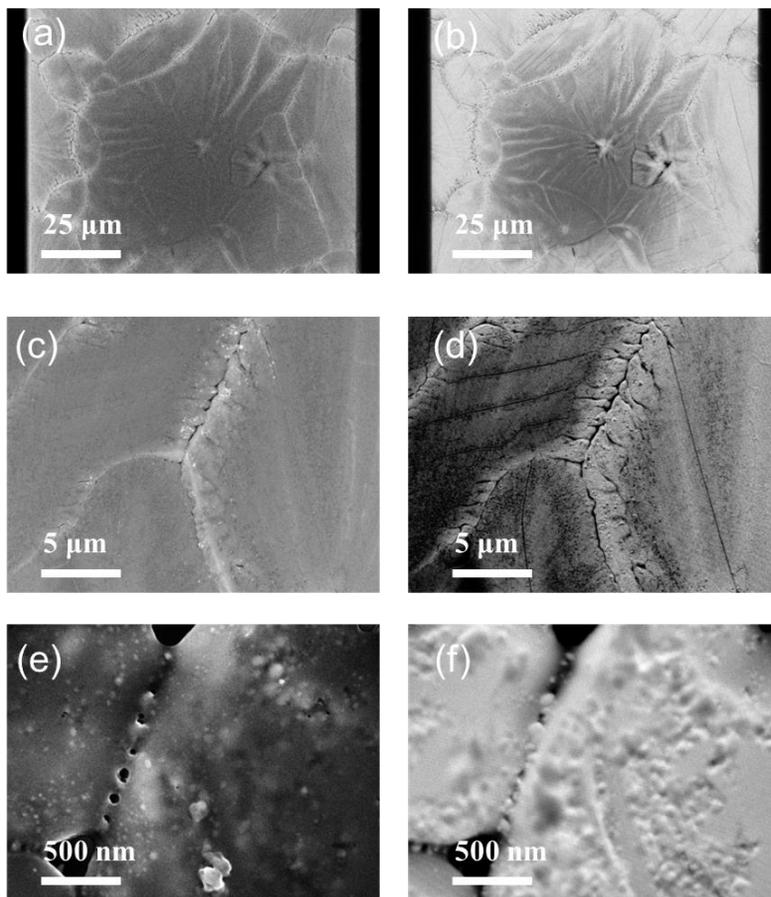

**Figure S22** (a, c, e) SEM (SE2) and (b, d, f) forescatter diodes in imaging for PDMAPb($I_{0.5}Br_{0.5}$)$_4$ thin film that had been de-mixed for 50 h at 1.5 mW/cm$^2$ at 80°C. The imaging was conducted in window area.

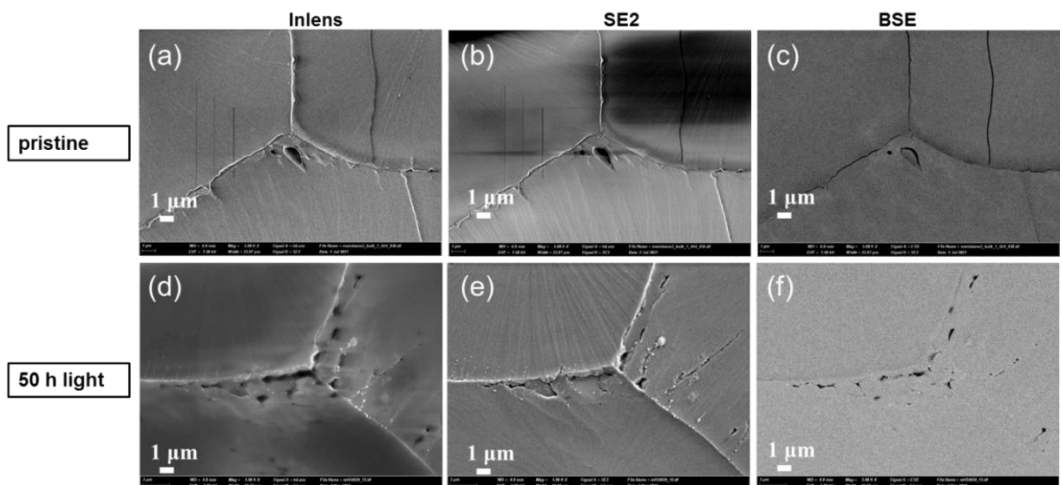

**Figure S23** SEM for PDMAPb($I_{0.5}Br_{0.5}$)$_4$ thin film. (a-c) pristine; (d-f) de-mixed for 50 h at 1.5 mW/cm$^2$ at 80°C. (a, d) InLens; (b, e) SE2; (c, f) BSE detectors. The imaging was conducted in non-window area. Magnification: 5K.

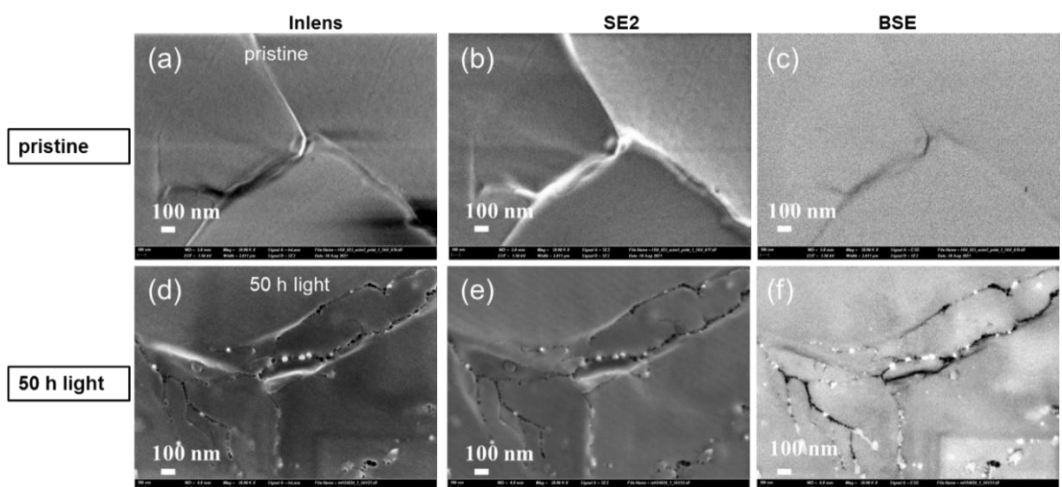

**Figure S24** SEM for a PDMAPb(I$_{0.5}$Br$_{0.5}$)$_4$ thin film. (a-c) pristine; (d-f) de-mixed for 50 h at 1.5 mW/cm$^2$ at 80°C. (a, d) InLens; (b, e) SE2; (c, f) BSE detectors. The imaging was conducted in non-window area. Magnification: 30K.

### S5.3 The nature of de-mixed domains

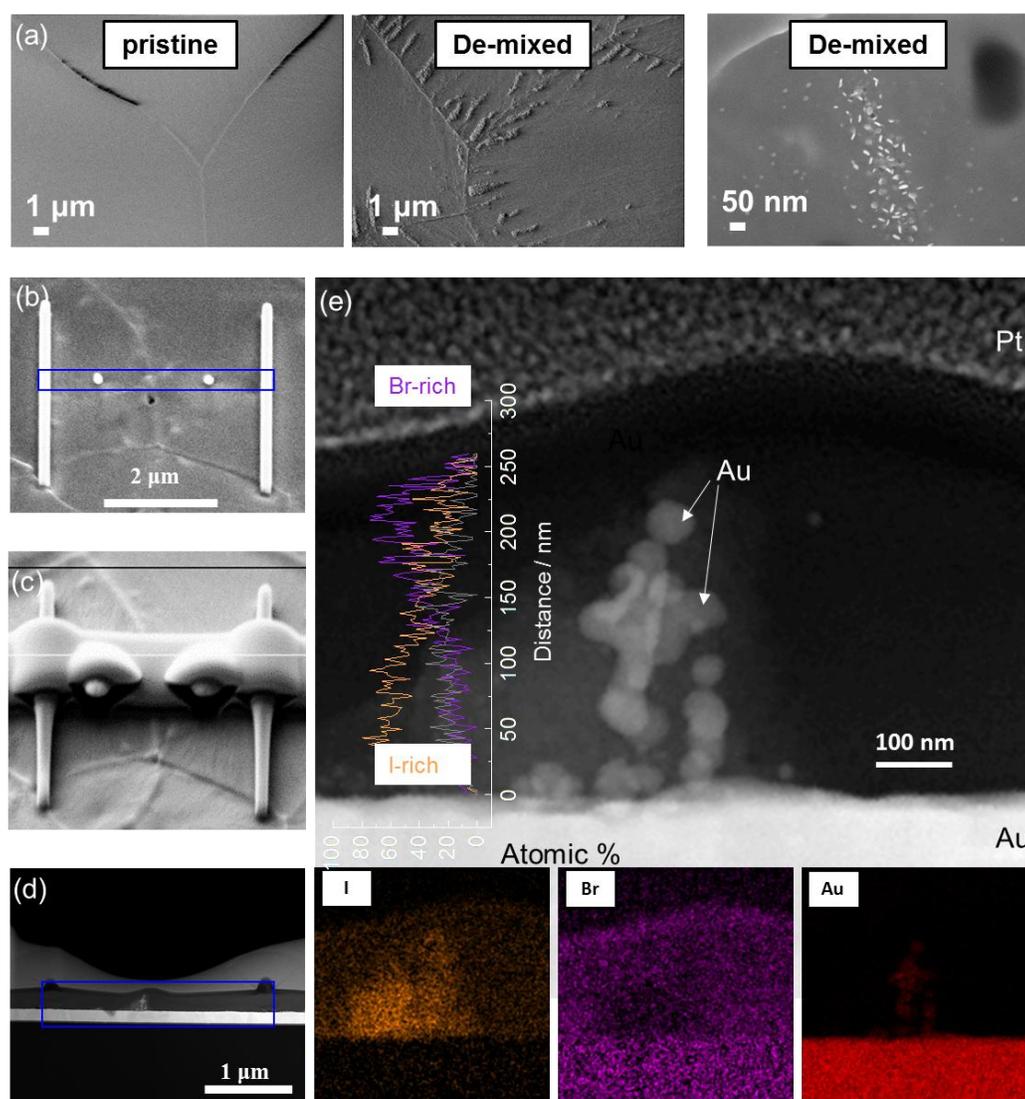

**Figure S25.** (a) SEM of a (PDMA)Pb(I$_{0.5}$Br$_{0.5}$)$_4$ thin film in pristine (left, magnification: 5K) and de-mixed state (illumination intensity 1.5 mW/cm$^2$ at 80°C, middle (magnification: 5K) and right (magnification: 50 K); Images from InLens detector are used. See complete data set with other detectors in Figure S20 and Figure S21. (b-e) Identification of the white crystallites formed after photo de-mixing under light (1.5 mW/cm$^2$) at 80 °C for 20 h (b) Top view SEM image after depositing carbon markers to identify the position of the target crystallite. The carbon pillars are used to mark the position of the grain boundary area and the carbon dots to mark the size of the crystallites for FIB lamella preparation. (c) Top view SEM image after depositing Pt. (d) TEM image for the area shown in (a), the black spots in this image are the cross sections of the carbon pillars in a.

(e) TEM image with higher resolution highlighting the crystallites on the grain boundary is I-rich domain (the bottom panel show the EDX mapping). Au nanoparticles are also found in the perovskite film region based on the TEM characterization, highlighting a more complex interaction between the film and gold contact during photo de-mixing.

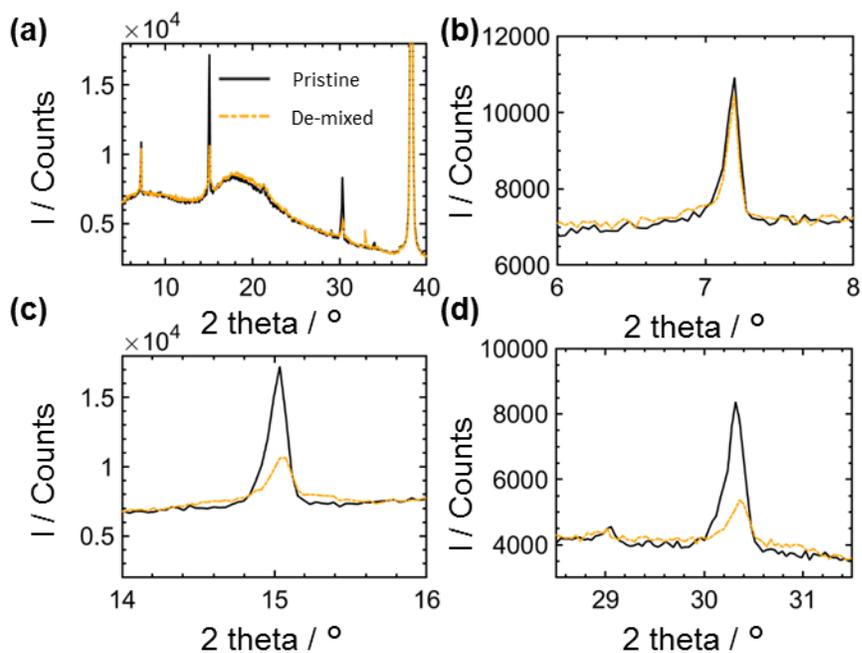

**Figure S26** XRD measurements of a (PDMA)Pb(I$_{0.5}$Br$_{0.5}$)$_4$ thin film (without PMMA encapsulation, in Ar atmosphere) before (black solid line) and after photo de-mixing (yellow dashed line, cold white LED illumination with 1.5 mW /cm² for 20 h at 80°C). (a) 2θ = 5°- 40°; The peaks are highlighted at (b) 2θ = 6°- 9°; (c) 2θ = 14°- 16°; (d) 2θ = 13°- 17°.

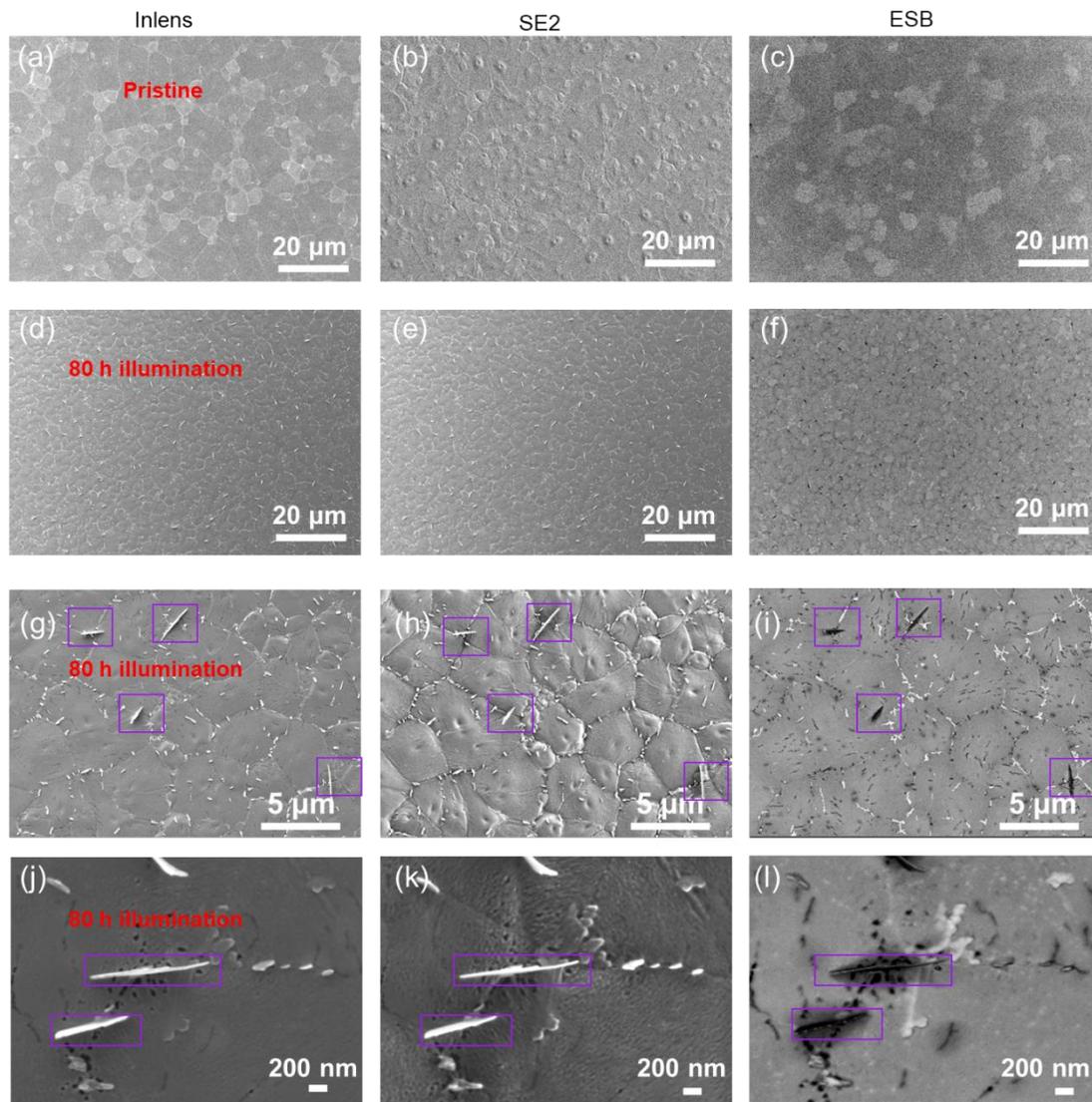

**Figure S27** SEM of a PDMAPb(I$_{0.5}$Br$_{0.5}$)$_4$ thin film without encapsulation after 80 h de-mixing time (illumination intensity 1.5 mW/cm$^2$ at 80°C). (a-c) pristine; (d-l) de-mixed for 80 h: (d-f) 1K; (g-i) 5K (j-l) 30 K. (a, d, g, j) InLens; (b, e, h, k) SE2; (c, f, i, l) BSE detectors. The perovskite thin film is deposited on a quartz substrate (also used for optical measurements).

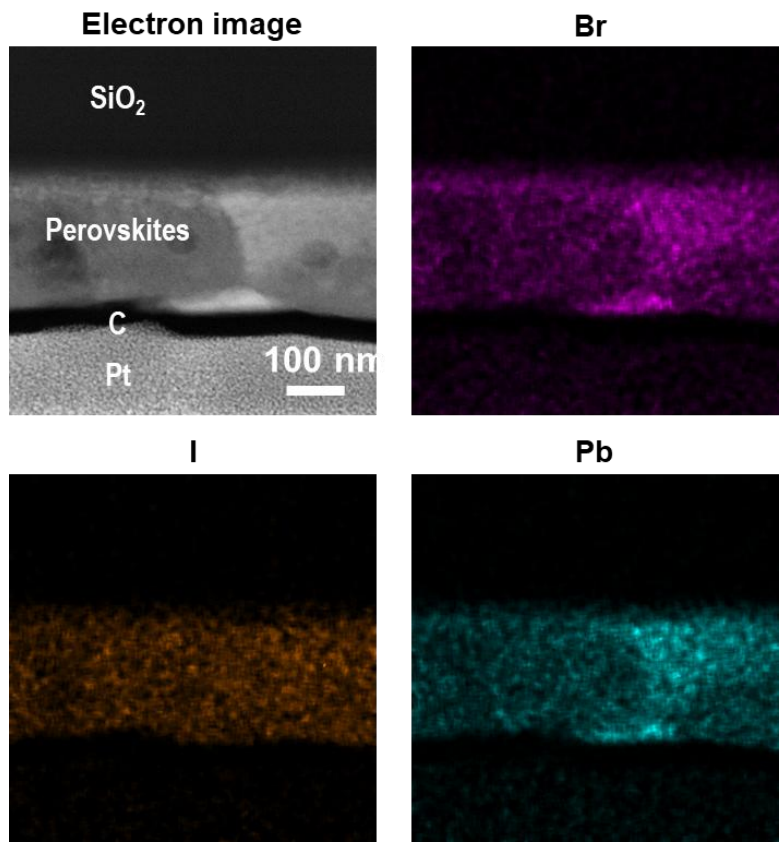

**Figure S28** TEM image and EDX mapping of a PDMAPb(I$_{0.5}$Br$_{0.5}$)$_4$ thin film (without encapsulation) after 80 h illumination (illumination intensity 1.5 mW/cm$^2$ at 80°C).

According to Table S12. Iodide to Pb ratio (I : Pb) and bromide to Pb ratio (Br : Pb) are 0.96 and 1.89 respectively, which suggests that both iodide and bromide can be excoporated from the bulk and leave the system, but iodine excoporation occurs in a larger extent. This is consistent with the observation from UV-Vis spectra that with longer time illumination that the peak corresponding to I-rich phase show a sharp decrease in absorption.

**Table S12.** EDX mapping on the area shown in Figure S28 The K factor is calibrated separately with PbI$_2$ and PbBr$_2$ stoichiometric power.

| Element | Extracted Spectrum Line Type | Extracted Spectrum Net Counts | Extracted Spectrum Net Counts err | Extracted Spectrum Atom % | Extracted Spectrum Atom % err | Extracted Spectrum K Factor |
|---|---|---|---|---|---|---|
| **Br K** | K | 19039 | 414 | 49.04 | 1.07 | 0.88 |
| **I L** | L | 26252 | 251 | 25.01 | 0.24 | 0.52 |
| **Pb L** | L | 23042 | 521 | 25.95 | 0.59 | 1.00 |
|  |  |  |  | 100.00 |  |  |

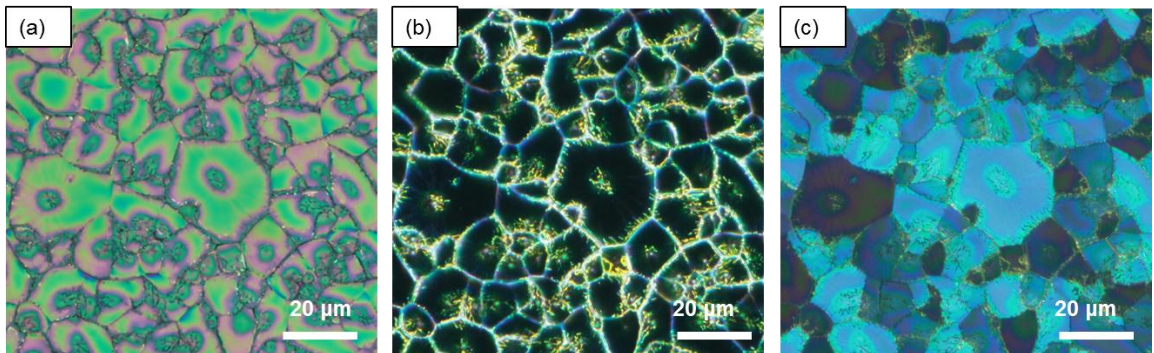

**Figure S29** (a) Bright field (b) dark field (c) polarized images of a (PDMA)Pb($I_{0.5}Br_{0.5}$)$_4$ thin film on a Si substrate.

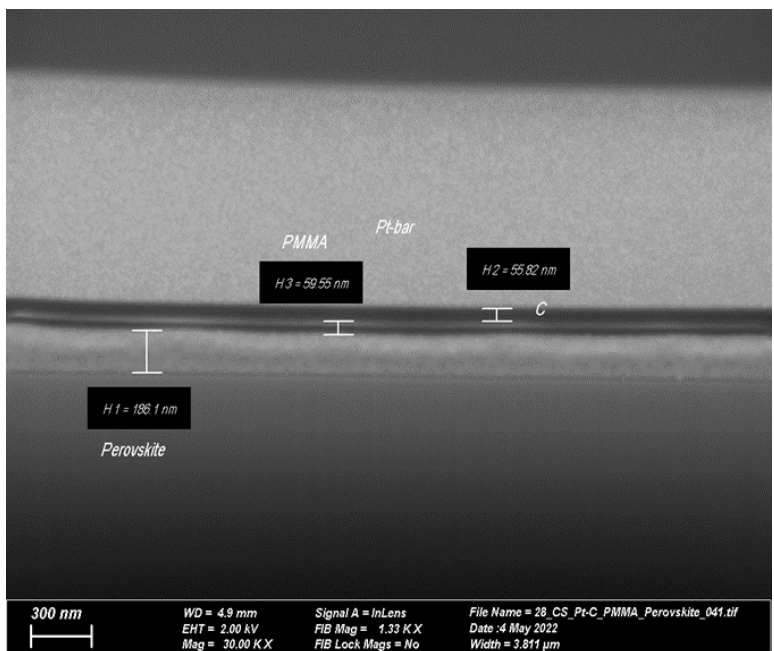

**Figure S30** Cross sectional image of a (PDMA)Pb($I_{0.5}Br_{0.5}$)$_4$ film on quartz substrates with PMMA encapsulation. From bottom to top, the layer are substrate (quartz), perovskites (186 nm), PMMA (60 nm), carbon and Pt-bar. The carbon and Pt bar are used for creating a conducting surface and protection layer for the cutting.

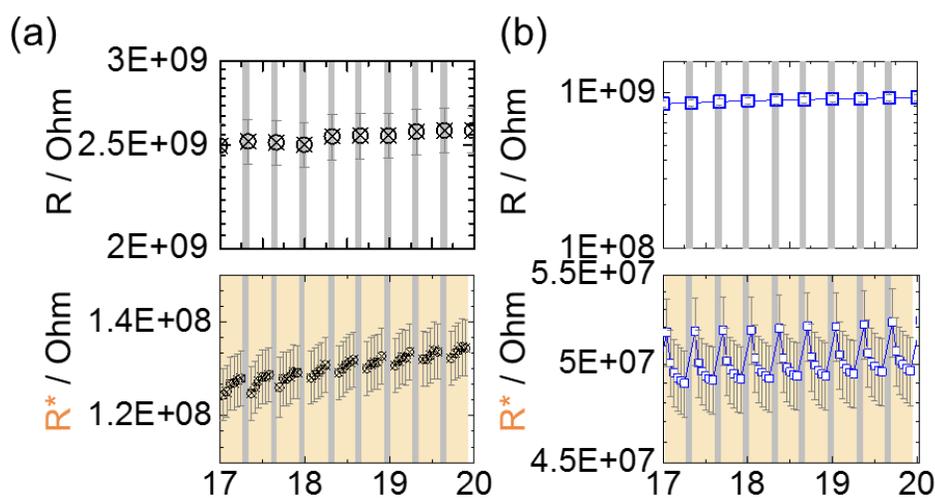

**Figure S31** Transient in electronic resistance (R* in yellow background) when quasi equilibrium of the photo de-mixing is reached. (a) Without PMMA encapsulation (b) with PMMA encapsulation on the surface prior to illumination. Ionic resistance remains almost constant in both case.

**S5.4 Partial reversibility of de-mixed domains**

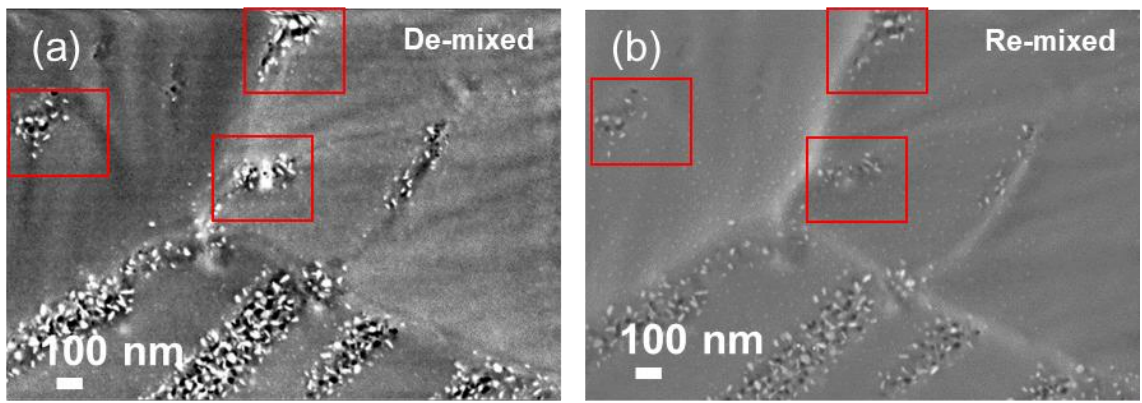

**Figure S32** SEM for (PDMA)Pb($I_{0.5}Br_{0.5}$)$_4$ thin film de-mixied for 50 h at 1.5 mW/cm$^2$ at 80°C (middle) and re-mixed in the dark in the glovebox at room temperature for 6 months. The red square indicates the disappearance of the crystallites after the dark storage treatment mentioned above. Images from SE2 detector are used. Magnification: 30K.

**S6 3D and NC mixed bromide – iodide perovskites**

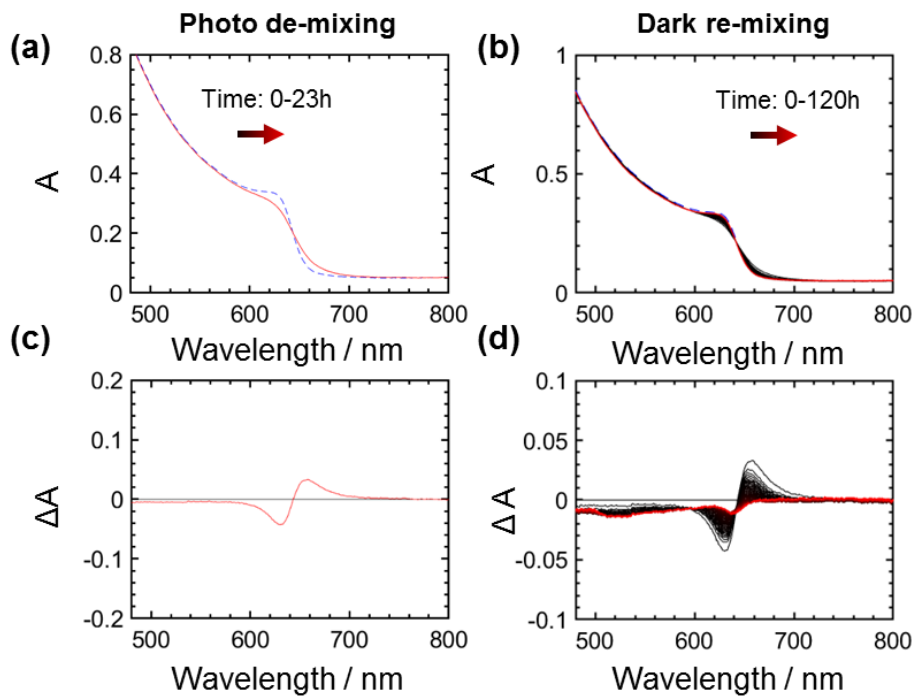

**Figure S33.** UV-Vis spectra evolution of MAPb($I_{0.5}Br_{0.5}$)$_3$ thin films (with PMMA encapsulation and in Ar atmosphere) (a) under light (1.5 mW/cm$^2$) at 40 °C for ~23 h with 20 min intervals (within each interval, spectra were collected by switching off the bias light for 300 s) and (b) in the dark at 40 °C for ~120 h with 1 h time interval between spectra. (c, d) Change in absorbance obtained by subtracting the reference spectrum of the pristine sample from each absorbance spectrum shown in (a) and (b), respectively.

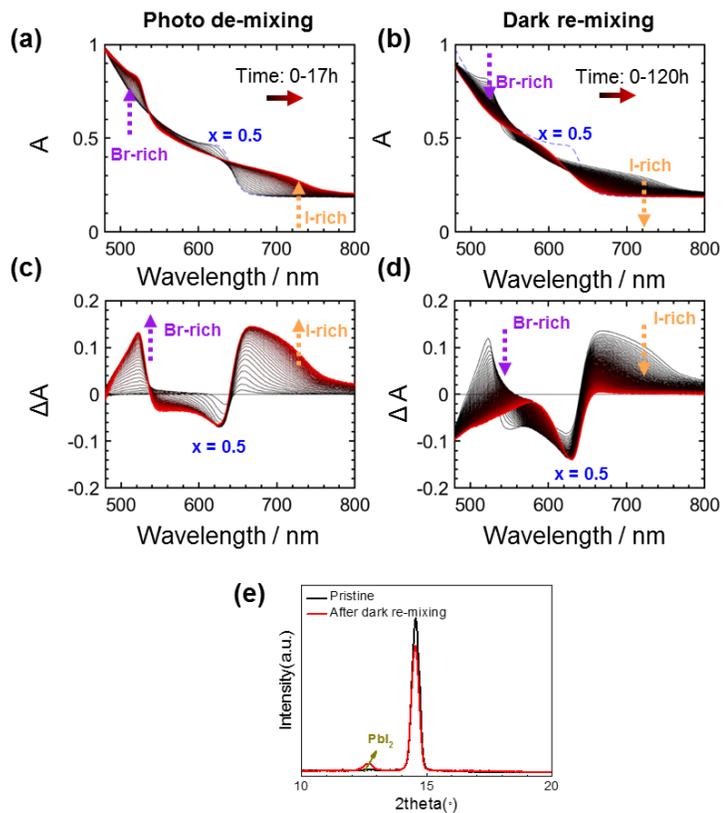

**Figure S34** UV-Vis spectra evolution of MAPb(I$_{0.5}$Br$_{0.5}$)$_3$ thin films (without PMMA encapsulation and in Ar atmosphere to exclude possible degradation) (a) under light (1.5 mW/cm$^2$) at 40 °C for 17 h and (b) in the dark at 40 °C for ~160 h with 1 h time interval between spectra. For the case under light, within each interval, spectra were collected by switching off the bias light for 300 s (see Figure 4.3 from the Material and Methods). (c, d) Change in absorbance obtained by subtracting the reference spectrum of the pristine sample from each absorbance spectrum shown in (a) and (b), respectively. (e) Change in X-ray diffraction patterns between the pristine state and the sample after dark re-mixing.

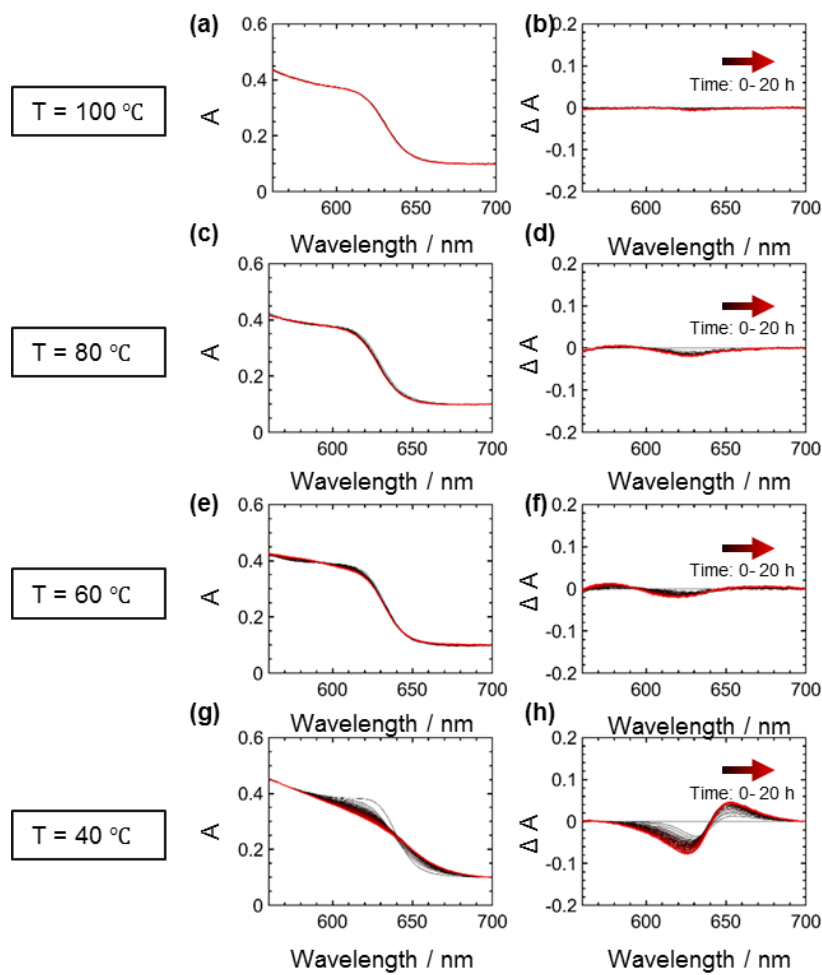

**Figure S35.** UV-Vis measurements performed on 3D MAPb(I$_{0.5}$Br$_{0.5}$)$_3$ thin films (with PMMA encapsulation) with different de-mixing temperature (T$_{de-mixing}$) at (a, b) 100 °C; (a, d) 80 °C; (e, f) 60 °C; (g, h) 40 °C.

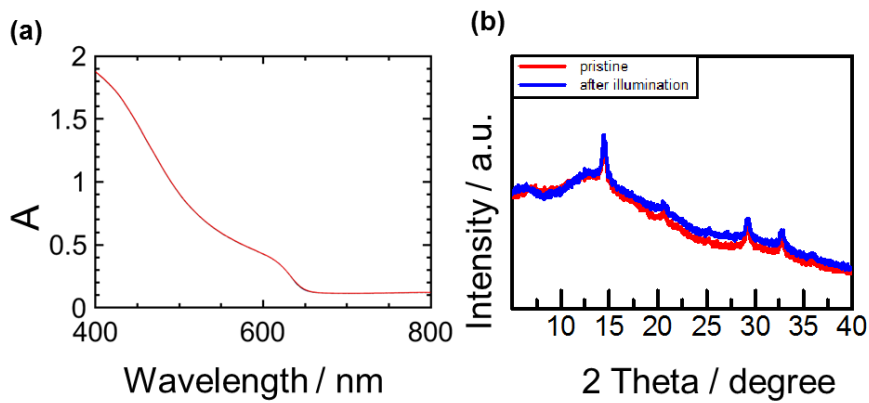

**Figure S36.** (a) UV-Vis spectra evolution of BA-MAPb(I$_{0.5}$Br$_{0.5}$)$_3$ nanocrystal thin films (without PMMA encapsulation and in Ar atmosphere to exclude possible degradation) (a) under light (1.5 mW / cm$^2$) at 40 °C for ~20 h with 20 min interval. (b) X-ray diffraction of the film before and after illumination.

**S7 Transport properties of nanocrystalline halide perovskite thin films**

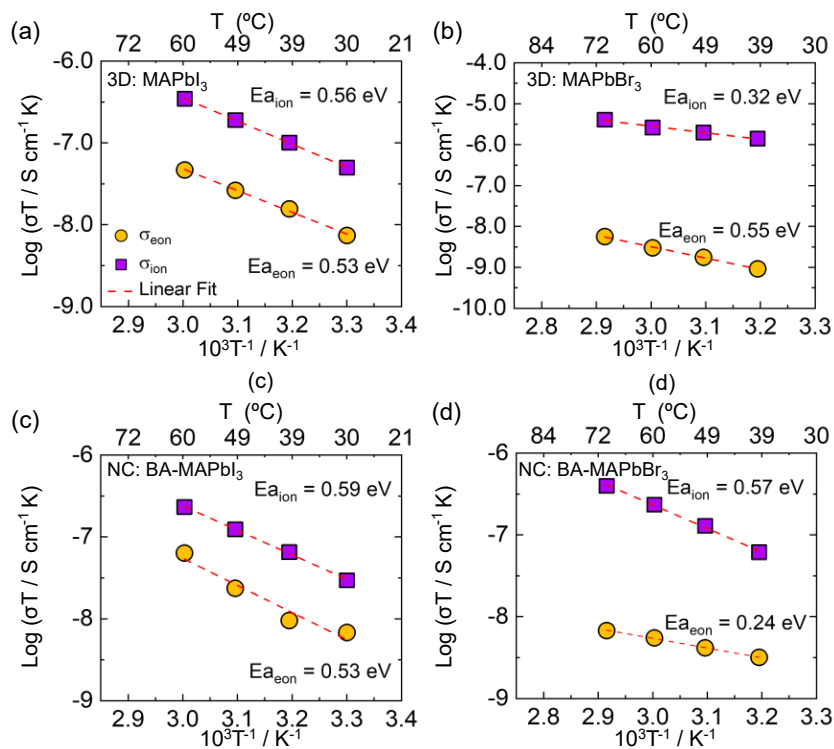

**Figure S37**. **The electronic conductivity ($\sigma_{eon}$, yellow circle) and ionic conductivity ($\sigma_{ion}$, purple square) of** in (a) 3D MAPbI$_3$ (b) 3D MAPbBr$_3$, (c) nanocrystalline BA-MAPbI$_3$ and (d) nanocrystalline BA-MAPbBr$_3$ thin films.

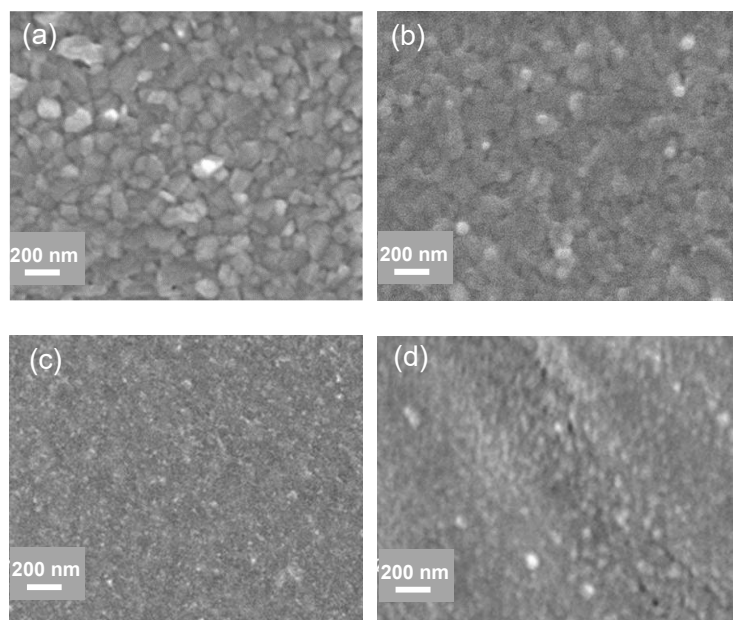

**Figure S38**. **SEM (SE2, secondary detector) images of** (a) 3D MAPbI$_3$ (b) 3D MAPbBr$_3$, (c) nanocrystalline BA-MAPbI$_3$ and (d) nanocrystalline BA-MAPbBr$_3$ thin films.